# Performance of trajectory surface hopping method in the treatment of ultrafast intersystem crossing dynamics

Jiawei Peng[1,2,4], Yu Xie[1], Deping Hu[3], and Zhenggang Lan[1,a)]

[1]*SCNU Environmental Research Institute, Guangdong Provincial Key Laboratory of Chemical Pollution and Environmental Safety & MOE Key Laboratory of Environmental Theoretical Chemistry, South China Normal University, Guangzhou 510006, China*

[2]*Qingdao Institute of Bioenergy and Bioprocess Technology, Chinese Academy of Sciences, Qingdao 2066101, China*

[3]*MOE Key Laboratory of Organic Optoelectronics and Molecular Engineering, Department of Chemistry, Tsinghua University, Beijing 100084, China*

[4]*University of Chinese Academy of Sciences, Beijing 100049, P. R. China*

We carried out extensive studies to examine the performance of the fewest-switches surface hopping method in the description of the ultrafast intersystem crossing dynamic of various singlet-triplet (S-T) models by comparison with the results of the exact full quantum dynamics. Different implementation details and some derivative approaches were examined. As expected, it is better to perform the trajectory surface hopping calculations in the spin-adiabatic representation or by the local diabatization approach, instead of in the spin-diabatic representation. The surface hopping method provides reasonable results for the short-time dynamics in the S-T model with weak spin-orbital coupling (diabatic coupling), although it does not perform well in the models with strong spin-orbital coupling (diabatic coupling). When the system accesses the S-T potential energy crossing with rather high kinetic energy, the trajectory surface hopping method tends to produce the good description of the nonadiabatic intersystem crossing dynamics. The impact of the decoherence correction on the performance of the trajectory surface hopping is system dependent. It improves the result accuracy in many cases, while its influence may also be minor for other cases.

---

[a)] zhenggang.lan@m.scnu.edu.cn

## I. Introduction

Nonadiabatic transitions widely occur within various molecular reactions, ranging from simple atomic/molecular collisions[1-3] to complicated photoinduced processes[4-7]. The understanding of the nonadiabatic process is an important challenge[4-22] due to the strong couplings between electrons and the nucleus. For example, different sophisticated methods[7, 10-16, 19, 20, 22-26] were developed for the simulation of nonadiabatic dynamics, while their employment to treat the extremely complicated systems with large numbers of degrees of freedom requires enormous computational costs. Practically, the employment of mixed quantum-classical approaches is always a promising approach to address the nonadiabatic dynamics of realistic complicated systems, by treating the nuclear motion with classical mechanics and the electronic motion in the quantum manner[4, 8, 23, 27-32]. This approach largely reduces the computational costs and has thus received considerable attentions in recent decades.

Over the past several years, many mixed quantum-classical methods have been developed to simulate nonadiabatic dynamic processes. Among them, the trajectory surface hopping (TSH) methods[4, 23, 28, 29, 33-42] are very popular in these methods, the classical trajectories are allowed to hop between different electronic states to describe the nonadiabatic transition. Particularly, the fewest-switches algorithm developed by Tully was widely employed[8, 28, 38, 43]. After the combination of the TSH methods and on-the-fly molecular dynamics, it is possible to describe the nonadiabatic dynamics of realistic polyatomic molecular systems at a fully atomic level with all degrees of freedom included[6, 38, 44-56]. Although Tully's TSH method and derivative approaches are widely developed and employed to study nonadiabatic dynamics, their deficiencies are also well known, such as the improper treatment of electronic coherence[5, 37, 38, 41, 42, 57-60]. Because it is not easy to derive these methods formally in a very rigorous fashion[61], it is still not easy to clearly judge the performance of Tully's TSH method in different cases, even after many theoretical developments and benchmark calculations[23, 28, 39-41, 54, 59, 62-77].

The TSH method is also very popular in the simulation of ultrafast intersystem crossing (ISC) processes[49, 78, 79], which involve the electronic transitions between the electronic states with different spin multiplicities. Various theoretical protocols were developed for an improved understanding of the ISC dynamics between different spin states. For instance, it is possible to compute the ISC dynamics by the TSH dynamics according to Landau-Zener[80] or Zhu-Nakamura theory[36]. In the early days, this idea was applied to study ISC relevant collision reactions[81-87]. Recently, the Zhu-Nakamura theory was combined with the on-the-fly simulation[88], which thus allows the simulation of nonadiabatic dynamics involving both internal conversion and ISC processes. When many electronic states are involved, it is also possible to apply Tully's fewest-switches surface hopping dynamics for the internal conversion dynamics between the same-spin states, while the ISC dynamics are treated by the Landau-Zener TSH approaches[89-91]. When the Tully's fewest switches TSH method is utilized to study ISC processes, it is possible to perform the dynamics calculations in the spin-diabatic representation[73, 91-97]. In this treatment, the eigenstates of the molecular Coulomb Hamiltonian are taken as the basis to represent the electronic wave function. Along this line, some works treat the internal conversion driven by the nonadiabatic coupling between the states with the same multiplicity and the intersystem crossing driven by the spin-orbital coupling (SOC) between the state with different spin multiplicities[91, 98]. In addition, Schatz, Maiti and coworkers once employed the mixed representation within the TSH treatment, in which the spin-adiabatic representation was employed to describe the potential energy surfaces in the reactant and product region, while the crossing region was described by the spin-diabatic representation[99-101]. Alternatively, the spin-adiabatic basis was recommended to elucidate the TSH dynamics[49, 73, 78]. The advantages of the TSH dynamics in the spin-adiabatic representation were discussed in the previous work[73]. To avoid some numerical instability problems, the local diabatization approach[102] (or similar approaches[103]) is an effective way to calculate TSH dynamics for the simulation of the ISC dynamics[73, 78, 104]. Currently the employment of the on-the-fly

TSH dynamics to study the ISC processes of realistic systems has also become routine, thanks to the developments of the on-the-fly TSH packages by several groups[49, 78, 97, 104, 105].

Several works once carefully checked the accuracy of the TSH dynamics[23, 28, 62-75, 106, 107]. Initialized by these important works, particular the interesting work by Persico and coworkers[73], we wish to evaluate the performance of Tully's TSH method in the simulation of the ultrafast ISC processes. Somehow, we wish that the ISC models used for the benchmark contain reasonable parameters belonging to the realistic value range. Starting from the multistate multimode model developed by Daniel, Gindensperger and coworkers in the treatment of a rhenium (I) tricarbonyl complex system ([Re(Br)(CO)$_3$bpy]; bpy = 2,2'-bipyridine)[108], we simplified such a model, choosing only one singlet state and a set of triplet states. This provides the linear vibronic Hamiltonian with four electronic states, from which we modify the relevant parameters to define several different testing models. The accuracy of the TSH method is examined by the comparison with the accurate results obtained from full quantum dynamics; namely, multiconfigurational time-dependent Hartree[12] (MCTDH) calculations. We attempted to verify different implementation details and TSH-relevant derivative approaches such as TSH in the diabatic or adiabatic representations, the local diabatization approach, etc. This work provides some useful complementary guidelines for the employment of the TSH method with respect to the ISC dynamics.

The rest of this paper is organized as follows: the Hamiltonian model and theoretical methods are briefly reviewed in Section II, including the four-state four-mode model Hamiltonian, the TSH method and the MCTDH method. The results and discussion are included in Section III. Conclusions are summarized in Section IV.

## II. Model Hamiltonian and Theoretical Methods
## A. The Simplified S-T Model

The linear vibronic coupling model that describes the excited-state dynamics of [Re(Br)(CO)$_3$bpy] is simplified to access the performance of the TSH method. As a typical halide transition-metal complex, [Re(Br)(CO)$_3$bpy] exhibits unique thermal and photochemical properties, and thus [Re(Br)(CO)$_3$bpy] and derivatives may potentially be used as molecular devices in various fields, such as sensors, probes, and imaging agents[108]. In addition, these types of systems display very interesting and complicated photophysical and photochemical processes[108-111], including the involvement of metal to-ligand charge transfer, internal conversion between different excited states, intersystem crossing to several high-spin states and strong dependence of electronic properties on local environments. [Re(Br)(CO)$_3$bpy] thus provides a prototype system for the study of photochemistry and photophysics of transition metal complex systems. In recent years, Daniel, Gindensperger and coworkers have performed substantial work in this area[108, 112-114], so we choose their model as our benchmark prototype. In their original paper[108], a six-mode five-state model including two singlet states ($S_1$, $S_2$) and three sets of triplet states ($T_1$, $T_2$, $T_3$) was constructed for the full-quantum MCTDH dynamics simulations. Based on this model a very complicated reaction mechanism was discussed, which includes the interplay between intersystem crossings and internal conversions. To set up simplified models suitable for the current test, we reduced the complexity of the model and only focused on the single ISC processes. The simplified model including one singlet state $S_2$ and a set of triplet states $T_1$ was built as the prototype $S$-$T$ model. Because the coupling modes in the original Hamiltonian, which couple different singlet states, should not play any role in the current ISC dynamic, only four normal modes ($v_7$, $v_{11}$, $v_{13}$ and $v_{30}$) were retained. It is very important to point out that the purpose of the current work is not for the comprehensive understanding of the photophysics and photochemistry of [Re(Br)(CO)$_3$bpy]. Instead, we wish to study the performance of the TSH method with respect to the description of the ISC processes governed by the SOC Hamiltonian with realistic parameters.

**B. The Model Hamiltonian**

The diabatic model Hamiltonian is expressed as follows:

$$H^{dia} = T_{nuc} + \sum_{i,j} |i\rangle H^{i,j}_{el,dia} \langle j|, \tag{1}$$

where $T_{nuc}$ represents the kinetic energy of the nuclei. The diagonal elements of $H^{i,i}_{el,dia}$ are the energies of the diabatic states, or more precisely the spin-diabatic electronic states, while the off-diagonal elements $H^{i,j}_{el,dia}$ characterize the interstate couplings of different states, namely, the SOC in the current model. In the dimensionless normal coordinates, the total Hamiltonian of the S-T model becomes the following:

$$\mathbf{H}^{dia} = \begin{bmatrix} E^{T^{(a)}}_{(0)} + \sum_i \kappa_i^{T^{(a)}} Q_i & 0 & 0 & \eta_{T^{(a)}-S} \\ 0 & E^{T^{(b)}}_{(0)} + \sum_i \kappa_i^{T^{(b)}} Q_i & 0 & 0 \\ 0 & 0 & E^{T^{(c)}}_{(0)} + \sum_i \kappa_i^{T^{(c)}} Q_i & \eta_{T^{(c)}-S} \\ \eta^*_{T^{(a)}-S} & 0 & \eta^*_{T^{(c)}-S} & E^S_{(0)} + \sum_i \kappa_i^S Q_i \end{bmatrix} + \frac{1}{2}\sum_i \omega_i (P_i^2 + Q_i^2) \mathbf{I}, \tag{2}$$

where $Q_i$ and $P_i$ are the dimensionless normal coordinate and momentum of mode $i$ with associated frequency $\omega_i$ respectively. $E^m_{(0)}$ ($m = S$, $T^{(a)}$, $T^{(b)}$, $T^{(c)}$) is the energy of electronic state $m$ and $E^{T^{(a)}}_{(0)} = E^{T^{(b)}}_{(0)} = E^{T^{(c)}}_{(0)} = E^T_{(0)}$, and $\kappa_i^m$ is the first-order intrastate vibronic coupling of mode $i$ and $\kappa_i^{T^{(a)}} = \kappa_i^{T^{(b)}} = \kappa_i^{T^{(c)}} = \kappa_i^T$. $\eta_{nS}$ ($n = T^{(a)}$, $T^{(c)}$) is the interstate coupling between $n$ and $S$ and $\eta_{T^{(a)}-S} = \eta^*_{T^{(c)}-S}$, $\mathbf{I}$ is a $4 \times 4$ unit matrix and the asterisk denotes the conjugate. The values of these parameters are listed in Table 1.

**Table 1. List of the parameters in the Model Hamiltonian. All values are reference values[108].**

| Mode | $\omega$ / eV | $\kappa$ / eV | |
|---|---|---|---|
| | | S | T |
| $v_7$ | 0.0116 | 0.0187 | -0.0161 |
| $v_{11}$ | 0.0188 | 0.0091 | 0.0002 |
| $v_{13}$ | 0.0229 | -0.0271 | -0.0261 |
| $v_{30}$ | 0.0792 | 0.0404 | -0.0196 |
| | $\eta$ / eV | | |

| | | |
|---|---|---|
| $T^{(a)}$-$S$ | | -0.0719 - 0.0196i |

Because of the large $S_2$-$T_1$ energy gap in the original model, the *S-T* crossing is far away from the Frank-Condon area. Thus, we need to modify some parameters to redefine several models suitable for the nonadiabatic ISC dynamics study. Four types of new models (I-IV) are constructed, in which the diabatic potential energy of the *S* state is fixed, while the *T* state is vertically shifted by adjusting $E_{(0)}^T$ until the *S-T* crossing appears not far from the Frank-Condon region. Because only one *S* state and a set of *T* states are involved, we will call all of the models below the *S-T* models in the following discussion. The relevant parameters in Models I-IV are collected in Table 2, and Figure 1 shows the diabatic potential energy curves along $Q_7$ (with the largest $|\kappa_i^T - \kappa_i^S|/\omega_i$). In Model I, the *S-T* crossing point is located exactly at the *S* minimum. In Models II and III, the *S-T* crossing points are on the right and left side of the *S* minimum energy point, respectively, and the potential energies of two points are equal to the energy of the first vibrational level (n($v_7$) = 1) along $Q_7$. In Model IV, the energy of the *S-T* crossing point is equal to the energy of the second vibrational level (n($v_7$) = 2) along $Q_7$, which is located on the right side of the *S* minimum.

**Table 2. List of transition energy values for Models I-IV**

| State | S | T | | | |
|---|---|---|---|---|---|
| Model | | I | II | III | IV |
| $E_{(0)}$ / eV | 3.1100 | 3.0539 | 3.1599 | 2.9479 | 3.2376 |

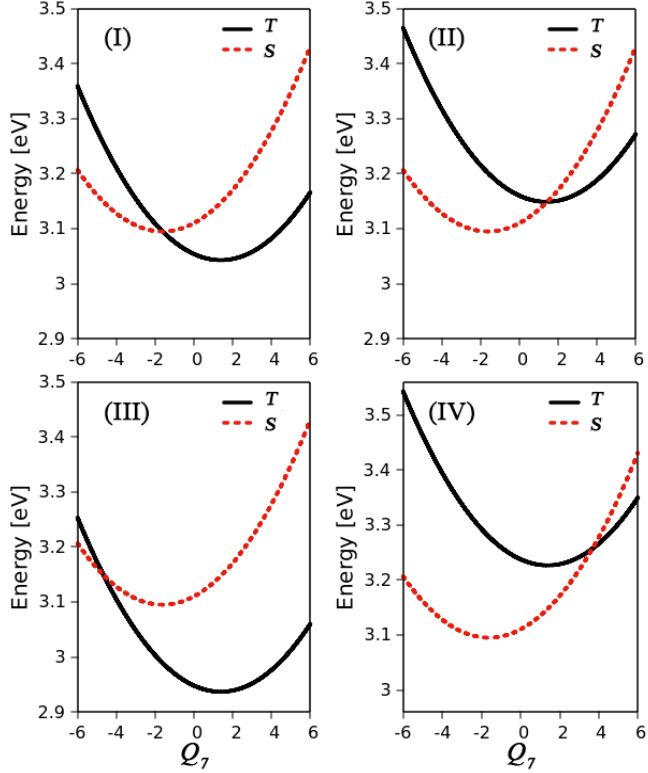

**Figure 1.** Diabatic potential energy surface (PES) along $Q_7$ for Models I-IV: the red dashed line denotes the PES of the singlet $S$ state and the black solid line indicates the PES of the triplet $T$ states.

As shown in previous work[73], it is possible to transform this type of $S$-$T$ model with one singlet state $S$ and a set of triplet states $T$ into a two-state model by employing the recombination of the electronic states, in which the singlet state is only coupled with a single triplet state, and the other two triplet states are completely decoupled. After the diabatic-to-adiabatic transformation, four adiabatic states are formed, which are labeled as $S_1^a$, $S_2^a$, $S_3^a$ and $S_4^a$ according to their energy order. Among them, the lowest state $S_1^a$ and the highest state $S_4^a$ are formed through the $S$-$T$ coupling, while the other two states $S_2^a$ and $S_3^a$ should be degenerated: their energies are the same as the original uncoupled triplet states.

When the combined triplet state basis is employed, only one effective triplet state is coupled with the singlet state. This provides a simplified method to perform the ISC dynamics. Instead of such a simplified way, we still execute dynamics in the original basis (Eq. (2)).

## C. The TSH Method

1. Theory

In Tully's TSH method[8], the nuclear degrees of freedom are treated by the classical method. Along trajectory propagation, the evolution of the total electronic wave function $\Phi(r,t;R)$ is governed by the time-dependent Schrödinger equation as follows:

$$H_e |\Phi(r,t;R)\rangle = i\hbar \frac{\partial}{\partial t} |\Phi(r,t;R)\rangle, \tag{3}$$

where $r$ is the electronic coordinate, $R$ is the time-dependent nuclear position, and $H_e$ is the electronic Hamiltonian parametrically dependent on $R$. Utilizing a set of complete electronic basis set, the electronic wave function can be expanded as follows:

$$\Phi(r,t;R) = \sum_i c_i(t) \varphi_i(r;R), \tag{4}$$

and $c_i(t)$ is the expansion coefficient corresponding to the electron wave function $\varphi_i(r;R)$. Therefore the time-dependent Schrödinger equation can be further reduced to the following:

$$i\hbar \frac{d}{dt} c_j(t) = \sum_i c_i(t) H_{ji} - i\hbar \sum_i c_i(t) d_{ji} \cdot v, \tag{5}$$

$$H_{ji} = \langle \varphi_j(r;R) | H_e | \varphi_i(r;R) \rangle, \tag{6}$$

$$d_{ji} = \langle \varphi_j(r;R) | \frac{\partial}{\partial R} | \varphi_i(r;R) \rangle, \tag{7}$$

where $v$ represent the vector of nuclear velocities, $H_{ji}$ is the element of the electronic Hamiltonian matrix, and $d_{ji}$ is the nonadiabatic derivative coupling vector. The above equation can be written in the density matrix framework as follows:

$$i\hbar \frac{d}{dt} \rho_{ij}(t) = \sum_k \left( H_{ik} \rho_{kj}(t) - \rho_{ik}(t) H_{kj} \right) - i\hbar \left( d_{ik} \rho_{kj}(t) - \rho_{ik}(t) d_{kj} \right) \cdot v, \tag{8}$$

where the electronic density matrix $\rho_{ij}$ is as follows:

$$\rho_{ij}(t) = c_i(t) c_j^*(t). \tag{9}$$

According to Tully's fewest switches algorithm, the hopping probability from state $i$

to state $j$ is evaluated by the following:

$$P_{ij}(t) = -\frac{2\int_{t}^{t+\Delta t} dt \left[ \hbar^{-1} \text{Im}\left(\rho_{ji}(t) H_{ij}\right) - \text{Re}\left(\rho_{ji}(t) \boldsymbol{d}_{ij} \cdot \boldsymbol{v}\right) \right]}{\rho_{ii}(t)}. \tag{10}$$

When the hopping probability is obtained, a random number is generated according to the uniform distribution over [0, 1]. The comparison of the random number and the hopping probability finally defines whether the trajectory hops to another state or not. If the hop takes place, the velocities are adjusted to confirm the total-energy conservation. In the case of frustrated hops, the hopping is rejected. In addition, the lack of decoherence[77] is one of the well-known problems in the standard TSH method. Thus, we employed the decoherence correction proposed by Persico and his coworkers[59] in some calculations, and the constant parameter $\alpha$ is set as 0.1 Hartree.

2. Representation

It is well known that TSH results are highly dependent on the representation[28, 39, 73, 74, 115, 116]. In principle, the following three different methods can be used to compute nonadiabatic dynamics.

(a) The first method, TSH-dia, only employs the diabatic representation for the propagation of both electronic and nuclear degrees of freedom. The hopping probabilities are computed in the same representation as well.

(b) The second approach, TSH-adi, involves the TSH method performed only in the adiabatic representation. In this case, the hops are governed by nonadiabatic couplings.

(c) In the third approach (local diabatization approach (TSH-loc) or other similar methods[49, 73, 78, 102, 103]), the nuclear motion is propagated in the adiabatic representation. The electronic motion is propagated in the diabatic representation or by the local diabatization approach, while the hopping probability is calculated in the adiabatic representation with the assistance of the diabatic-to-adiabatic transformation.

It is important to notice that the TSH-loc and TSH-adi approaches essentially employ the same representation. Both use the adiabatic representation in nuclear propagation, while they primarily differ by the algorithm used to evaluate the nonadiabatic couplings. The TSH-adi approach attempts to calculate the nonadiabatic coupling directly, while the TSH-loc approach evaluates the dot product of the nonadiabatic vector and velocity indirectly by using the diabatic-to-adiabatic transformation of relevant time-dependent physical quantities in the diabatic representation. Thus, in principle, these two approaches should converge to the same result in the same suitable time step. In the current work, all three approaches are taken into account for comparison. In the TSH-adi and TSH-loc approaches, the decoherence correction is added into the adiabatic representation.

3. Initial Conditions

The current work considers that the system initially stays on the electronic ground state minimum, thus the lowest vibrational level is taken to perform the initial sampling of the nuclear degrees of freedom. The classical action-angle variables are used to obtain initial coordinates and momenta[4, 70]. $Q_i$ and $P_i$ of the dimensionless normal modes are given by the following:

$$Q_i = \sqrt{2n_i + 1} \sin \alpha_i, \tag{11}$$

$$P_i = \sqrt{2n_i + 1} \cos \alpha_i, \tag{12}$$

where $\alpha_i$ represents the random angle over the range of [0, 2π], and $n_i$ is the corresponding quantum number of a harmonic oscillator; it is zero in this work.

At time zero, we vertically place these initial conditions into the singlet state $S$ in the spin-diabatic representation to initiate the trajectory propagation, according to the Condon approximation. Following the work by Müller and Stock[70], it is rather general to set the initial electronic coefficient of the singlet state S as $e^{i\theta}$ ($\theta$ is a random number picked from the interval [0,2π]), while these coefficients are zero for the triplet states $T$. In the spin-adiabatic representation, the initial electronic coefficients of the spin-adiabatic states are obtained from the diabatic-to-adiabatic transformation. For operational details, readers may refer to the previous works by Müller and Stock[70]. In fact, the introduction of the phase factor $\theta$ may result in only a negligible influence on the ISC dynamics in this work, because the electronic populations are not dependent on such a factor in the current calculations. In more general cases, for instance in the presence of a strong coherent laser field or with the involvement of many close-lying states, the initial phase may become important.

4. Adiabatic and Diabatic State Population and Occupation

In the TSH simulations, the state occupation is defined as the percentage of trajectories propagating on different potential energy surfaces as follows:

$$P_k^{occ}(t) = N_k(t) / N_{total}, \tag{13}$$

where $N_{total}$ represents the total number of trajectories and $N_k(t)$ is the number of trajectories on state $k$ at time $t$. The state population is calculated according to the average of the time-dependent electronic populations of the different states as follows:

$$P_k^{pop}(t) = \langle \rho_{kk}(t) \rangle_{traj}. \tag{14}$$

We borrow the above nomenclature of "state occupation" and "population" to distinguish Eq. (13) and Eq. (14) from the previous work[70], because this simplifies illustration.

If the TSH-dia method is employed, the diabatic state occupation and electronic population are directly gained from simulations, while the adiabatic electronic population is obtained by the diabatic-to-adiabatic transformation.

If the TSH-adi calculation is employed, the opposite method is true and two different methods can be used to calculate the corresponding diabatic population. The first method occurs directly by the adiabatic-to-diabatic transformation of the density matrix (via Eq. (15)) and then performs the average over all trajectories, namely:

$$P_i^{dia}(t) = \left\langle \left( \mathbf{U}^+ \boldsymbol{\rho}^{adi}(t) \mathbf{U} \right)_{ii} \right\rangle_{traj}, \tag{15}$$

where $\mathbf{U}$ is the relevant transformation matrix. We also consider the second method[40,]

[62, 63] to obtain the diabatic population. We reconstruct the effective adiabatic electronic density matrix by setting the diagonal element to 1 (or 0) and keep the off-diagonal element unchanged. This effective density matrix is then employed to perform the adiabatic-to-diabatic transformation and the average is calculated (via Eq. (16)) as follows:

$$P_i^{dia}(t) = \left\langle \left(\mathbf{U}^+ \boldsymbol{\rho}_{eff}^{adi}(t) \mathbf{U}\right)_{ii} \right\rangle_{traj} \left( \rho_{eff,ii}^{adi}(t) = \begin{cases} 1 & i = current\_state \\ 0 & i \neq current\_state \end{cases}, \rho_{eff,ij}^{adi}(t) = \rho_{ij}^{adi}(t) \right). \quad (16)$$

If the TSH-loc approach is employed, the diabatic electronic population is directly obtained from the calculations, and then the diabatic-to-adiabatic transformation yields the hopping probability. In this way, both the adiabatic population and the adiabatic state occupation are obtained automatically.

5. Numerical Details

To conduct the systematic benchmark calculations, several sets of TSH simulations are performed. In each case, 2500 trajectories are propagated within the surface-hopping dynamics. The evolutions of nuclei and electrons are integrated using the fourth-order Runge-Kutta method with different time steps. To produce converged results, the time step of nuclear motion is set as 0.1 fs, which is 100 times greater than the time steps of the propagation of the electronic motions, and the propagation finishes at 1ps. To check the dependence of the results on the time step, we also use different nuclear time steps or electronic time steps in some calculations; see below. Here, the time-dependent Schrödinger equation for electronic motion contains the potential energies, the dot product of velocities and nonadiabatic coupling. At each electronic time step, the coordinates and velocities of the nuclei are obtained by linear interpolation. The relevant energies, gradients and nonadiabatic coupling vectors are computed directly from the model Hamiltonian based on the interpolated position and velocity.

### D. The MCTDH Method

In the MCTDH method[12, 117, 118], the time-dependent basis is used to expend the wave function. For a system with $f$ degrees of freedom, the wave function $\Psi(Q_1,...,Q_f,t)$ is expressed as the following

$$\Psi(Q_1,...,Q_f,t) = \sum_{j_1=1}^{n_1}...\sum_{j_f=1}^{n_f} A_{j_1...j_f}(t) \prod_{\kappa=1}^{f} \varphi_{j_\kappa}^{(\kappa)}(Q_\kappa,t), \tag{17}$$

where $Q_1,...,Q_f$ represent nuclear coordinates, and $A_{j_1...j_f}$ are the time dependent expansion coefficients corresponding to the time-dependent single particle function $\varphi_{j_\kappa}^{(\kappa)}$. Utilizing the variational principle, the coupled equations of motion can be given by the following:

$$i\frac{\partial}{\partial t} A_{j_1...j_f} = \sum_{l_1...l_f} \left\langle \prod_{\kappa=1}^{f} \varphi_{j_\kappa}^{(\kappa)} \middle| H \middle| \prod_{\kappa=1}^{f} \varphi_{l_\kappa}^{(\kappa)} \right\rangle A_{l_1...l_f}, \tag{18}$$

$$i\dot{\varphi}_m^{(\kappa)} = \left(1 - P^{(\kappa)}\right) \sum_{j,l=1}^{n_\kappa} \left[\left(\boldsymbol{\rho}^{(\kappa)}\right)^{-1}\right]_{mj} \langle H \rangle_{jl}^{(\kappa)} \varphi_l^{(\kappa)}, \tag{19}$$

where $P^{(\kappa)}$ represents the projection onto the space spanned by the single-particle functions for the $\kappa$th degree of freedom, $\langle H \rangle_{jl}^{(\kappa)}$ is the relevant mean-field Hamiltonian acting on the $\kappa$th degree of freedom, and $\left(\boldsymbol{\rho}^{(\kappa)}\right)^{-1}$ is the inverse matrix of density for the $\kappa$th degree of freedom.

In the current MCTDH calculations, the initial wave packet is obtained by vertical excitation of the ground vibrational level of the electronic ground state to the $S$ state.

The diabatic population $P_\alpha^{dia}(t)$ is defined by the expectation value of the projector $|\psi_\alpha^{dia}\rangle\langle\psi_\alpha^{dia}|$ with the time-dependent wave function $|\Psi(t)\rangle$

$$P_\alpha^{dia}(t) = \langle\Psi(t)|\psi_\alpha^{dia}\rangle\langle\psi_\alpha^{dia}|\Psi(t)\rangle, \tag{20}$$

where $|\psi_\alpha^{dia}\rangle$ is the diabatic basis.

### III. Results and Discussion

### A. Representations

We first consider the performance of the TSH-dia approach. Figure 2 shows the results of the time-dependent population dynamics. For easy comparison, only the diabatic population of the $S$ state is plotted here.

In Eq. (2), one of the triplet states does not couple with other electronic states in the current model Hamiltonian, and thus its population remains unchanged (see Appendix I). The population of the two other triplet states remains the same due to the same coupling strength, although their SOCs display different phases.

In all four models, the accurate quantum dynamic MCTDH results exhibit extremely fast oscillations of the diabatic population, which means that the SOCs between different spin-diabatic states are very strong. Such rapid oscillation is relevant to the pure electronic motion and very similar to Rabi-type oscillation. The oscillation amplitude itself also displays the long-period modulated pattern, and this long period is consistent with the one associated with the slowest vibrational motion of the $v_7$ mode.

The red dotted and green full lines show the state occupation and the electronic population of the spin-diabatic $S$ state obtained with the TSH-dia approach, respectively. Although the oscillation of the electronic population is observed as well, such oscillation not only decays much faster, but its amplitude is also much smaller than the exact result. The state occupation, on the other hand, yields even worse results, and almost no oscillation is observed. Therefore, in these models, the TSH-dia approach does not provide reasonable dynamics results. The reasons for this may be possibly attributed to the following: when the trajectory propagates in the domain with very small diabatic energy gap, many hops may happen between different spin-diabatic states. As a result, such a treatment may not fully satisfy the essential idea of the "fewest-switches" assumption.

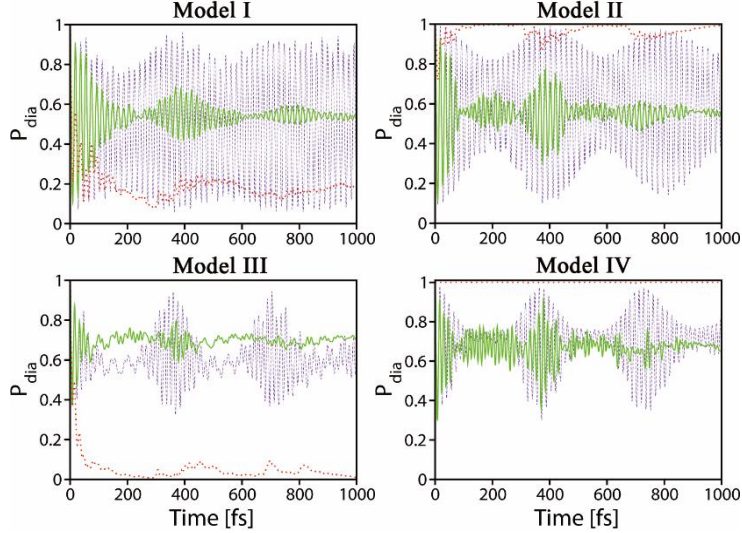

**Figure 2.** Time-dependent diabatic population dynamics: the diabatic electronic population of the MCTDH result (violet dashed line), the diabatic state occupation (red dotted line) and the diabatic electronic population (green full line) of the TSH-dia result.

When the TSH-adi or TSH-loc calculations are performed, the time-dependent evolution of the spin-adiabatic state with the largest initial population is given in Figure 3.

As discussed in the previous section (II.B), the effective model in the adiabatic representation only includes two coupled spin-adiabatic states. For two uncoupled spin-adiabatic states, their populations remain unchanged with respect to dynamics (see Appendix I). In Model I (Figure 3(a1)), it is observed that the population of the spin-adiabatic state with the largest initial population almost remains constant in the MCTDH calculations, which can be reproduced by the electronic population (Eq. (14)) in the TSH-adi dynamics. In the current model with strong SOC, the nonadiabatic coupling elements should be quite small due to the rather large energy gap. Thus, this term can be nearly neglected in the time-dependent Schrödinger equation no matter whether the full quantum version or the mixed quantum classical version is employed. Therefore, the adiabatic electronic population always remains unchanged in both the MCTDH result and the TSH-adi adiabatic electronic population result. As a contrast, the TSH-adi approach yields obvious and faster decay in the state occupation (Eq. (13)) with time. It is interesting to note that this occupation decay is almost monotonic, and that it finally reaches a stable value with time, which means that the trajectory remains on the lower spin-adiabatic state after hops and does not jump back to the original higher spin-adiabatic state. A similar situation exists for Model III because the upper spin-adiabatic state exhibits the larger initial population. This finding is also partially attributed to the large number of frustrated hops after trajectory jumping to the lower state (see Appendix II).

For Models II (Figure 3(a3)) and IV (Figure 3(a4)), the MCTDH population and the TSH-adi electronic population are consistent, while the TSH-adi calculations predict different state occupation evolution. This is very similar to the cases of Models I and III. The only difference is that the state occupation increases with the time evolution because the spin-adiabatic state with the large initial population shows lower energy in Models II and IV.

When the TSH-loc approach is taken, we obtain almost identical results with respect to the results in the TSH-adi calculations. This means that the two approaches, TSH-adi and TSH-loc, yield almost equivalent results in these models.

When the decoherence correction is included in the TSH-adi calculations, the state occupation and electronic population become identical, while both of them show deviation with respect to the MCTDH result.

For Model I-IV, the TSH-adi or TSH-loc approaches may capture the partial dynamic feature, while the TSH-dia approach may yield results which are far from accurate. These findings are consistent with the conclusions of previous TSH works[28, 39, 74, 115, 116]. For example, in the treatment of the one-dimensional scattering problem[28], Tully showed that the TSH results are dependent on the representation and that the diabatic representation is not normally a good choice. In the later part of this paper, we will provide more discussion on this topic.

It is clear that the TSH method does not work well in Models I-IV, possibly relevant to the strong SOC, no matter which representation is used for propagation. In Models I-IV, the TSH-dia approach is not trustable at all, while the TSH-adi and TSH-loc approaches may overestimate the decay from the upper state to the lower state. In the adiabatic representation, such situations correspond to the models with large adiabatic energy gaps and small nonadiabatic couplings. In other words, when the dynamics become more "adiabatic" and the nonadiabatic dynamics are essentially missing, the TSH method does not perform very well. In this situation, the nonadiabatic transition from the upper state to the lower state is always overestimated, while the reverse transition is strongly underestimated, particularly under the existence of the frustrated hops.

In Models I-IV, the existence of a large number of frustrated hops certainly destroys the reliability of the TSH-adi approach. However, the unsatisfactory performance of the TSH-adi approach is also associated with other reasons in addition to frustrated hops. It is well known that Tully's surface-hopping algorithm results in overcoherent electronic motion for a single trajectory. In these models, the overcoherent electronic motion of a single trajectory displays the fast Rabi-type oscillation of the electronic population due to the existence of the weak nonadiabatic couplings and the visible energy gap (see Appendix III). Such rapid Rabi-type oscillation may result in hops, when the sudden change of the electronic population occurs. As a consequence, the occupation decay from the upper state to the lower state is significantly overestimated. With the existence of frustrated hops that prevent the system from jumping back from the lower state to the upper one, the situation worsens. In this sense, the unsatisfactory results of the TSH-adi approach in Models I-IV are relevant to the nonphysical description of the electronic coherence, instead of only frustrated hops. Thus, when the decoherence correction is included, the performance of the TSH-adi approach improves (Figure 3(c1-c4)). At the same time, the number of frustrated hops may also be largely suppressed, also resulting in better results (see Appendix II). However, the decoherence correction does not give the fully satisfied description of the electronic motion, thus the results are still not exact. This strongly indicates that the TSH-adi method may not represent a good approach for the description of such high-frequency Rabi-oscillation features of the electronic evolution in the nuclear-electronic coupled dynamics, when the large energy gap and the weak nonadiabatic coupling appear in the models under study. This produces the key reason to explain the unreasonable performance of the TSH-adi approach in the model with weak nonadiabatic coupling and visible energy gap. In fact, Granucci and Persico once discussed this finding more comprehensively, and interested readers can check their work for more detailed discussions of this issue[59].

To further explore the performance of the TSH method, we changed the parameters of Hamiltonian and simulation details such as vibronic coupling and initial preparation. Next, we only consider changing parameters in Model III.

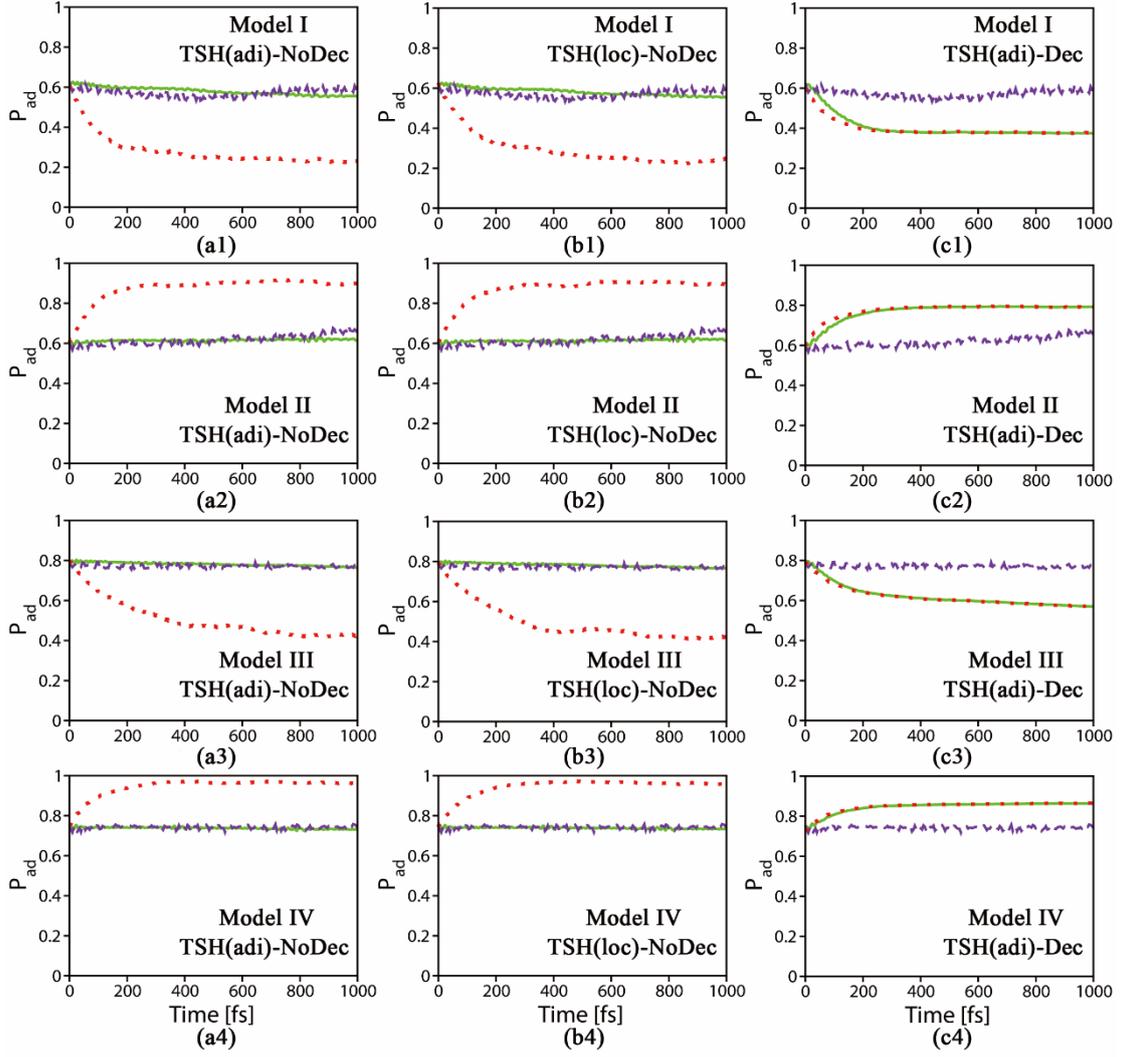

**Figure 3.** Time-dependent adiabatic population dynamics: the adiabatic electronic population of the MCTDH result (violet dashed line), the adiabatic state occupation (red dotted line) and the adiabatic electronic population (green full line) of the TSH results. The models employed are given in each figure. The label "adi" denotes the TSH-adi approach. The label "loc" denotes the TSH-loc approach. The label "Dec" (or "NoDec") denotes that the decoherence correction is used (or not).

## B. Spin Orbital Coupling

Starting from Model III, we rescaled the SOC by the factors of $\lambda$ = 0.15, 0.25 and 0.5 to give a series of models. The MCTDH and TSH results are given in Figure 4.

When the SOC (diabatic) weakens, the *S-T* state mixing lessens, resulting in the significant increase of the initial electronic population of the upper spin-adiabatic state. In this case, the decay of the population of the upper spin-adiabatic state becomes pronounced, and population recurrences are observed in the MCTDH calculations. When the TSH-adi calculations are performed, the adiabatic electronic population (Eq. (14)) reproduces the short-time dynamics and does not display the correct long-time behavior. In contrast, the adiabatic state occupation (Eq. (13)) now produces much better results, particular in the weak SOC cases, although it still does not reproduce the population recurrence. The lack of recurrence in the population dynamics was noticed in the previous treatments of other models, such as conical intersection[69, 70, 106] and the model of the harmonic potentials spanned by two coupled coordinates, namely, the heavier and lighter particle coordinates[75].

The effective performance of the TSH-adi dynamics may be attributed to the following fact: in the weak spin-diabatic coupling cases, the minimum energy gap between two spin-adiabatic states may become very small, leading to a very "local" avoided potential energy crossing in the adiabatic representation. This implies that the hops may easily occur in these crossing regions with small energy gaps. At the same time, the small energy gap may also reduce the possibility of frustrated hops. We also noticed that Tully and coworkers[75] once discussed the general performance of the TSH method. In their view, the TSH method may lead to improper treatment when the interstate coupling is highly delocalized. The scattering problem corresponds to the situation in which the model only contains the coupling highly localized in the interaction region. For the situations with delocalized couplings, two problems may cause the unsatisfactory results of the TSH-adi approach, namely, the nonphysical loss of coherence and the existence of many frustrated hops. In our current *S-T* model, we can also understand the TSH-adi performance within such a framework. In the weak SOC model, the energy gap is very small at the crossing in the spin-adiabatic representation, leading to quite localized nonadiabatic coupling. The small energy gap also reduces the number of frustrated hops. Therefore, the TSH-adi performance is enhanced.

When the decoherence correction is included, the state occupation and the electronic population become very similar. Both of them are comparable with the MCTDH results, particular for the weak coupling cases, while the population recurrence in the MCTDH dynamics is apparently still not captured by the TSH-adi dynamics with decoherence correction.

We also showed the diabatic population in Figure 5. When the TSH-adi dynamics are chosen, the diabatic population may be trivially obtained from the direct adiabatic-to-diabatic transformation via Eq. (15), and the relevant results are shown in Figure 5(a1-a3). Alternatively, it is also possible to utilize Eq. (16) to calculate the diabatic populations, as shown in Figure 5(b1-b3). It is interesting to see that the latter approach seems to work better and produces excellent short-time dynamics for the weak SOC (small diabatic coupling) cases. This means that the latter approach should represent a reasonable way to calculate the diabatic population. We noticed that such a method is also recommended by previous works[40, 62, 63]. When the decoherence correction is added (see Figure 5(c1-c3)), two approaches (Eq. (15) and Eq. (16)) yield very similar results with respect to the calculations of the diabatic populations. Similar to the discussion of the adiabatic population, the diabatic population also seems to show that the TSH-adi method works well in the weak SOC (weak diabatic coupling) cases.

When the TSH-loc method was employed, we obtained very similar dynamics as compared with the TSH-adi dynamics. In the TSH-loc approaches, the decoherence correction was added in the spin-adiabatic representation after the adiabatic coefficients were obtained from the diabatic-to-adiabatic transformation. The reversed transformation then provides the diabatic population under the decoherence correction (Figure 5(d1-d3)), and this approach produces results which are consistent with the ones obtained by the TSH-adi approach with decoherence correction.

Overall, the TSH methods (TSH-adi and TSH-loc) seem to yield reasonable results for the short-time dynamics when the weak SOC is presented. The inclusion of the decoherence correction improves the internal consistency between the state occupation and the population. When we calculate the diabatic populations in the TSH-adi approach, it is always effective to compute them via Eq. (16).

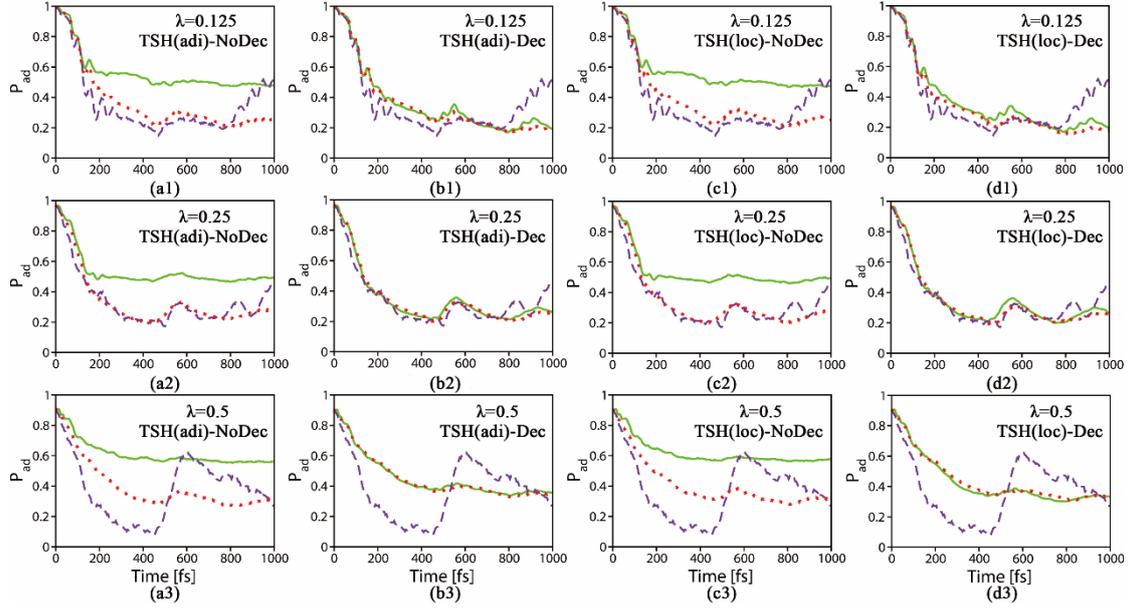

**Figure 4.** Time-dependent adiabatic population dynamics at different couplings: the adiabatic electronic population of the MCTDH result (violet dashed line), the state occupation (red dotted line) and the electronic population (green full line) of the TSH result. Starting from Model III, we rescale the SOC; the scaling factor λ is given in each subfigure. The label "adi" denotes the TSH-adi approaches. The label "loc" denotes the TSH-loc approach. The label "Dec" (or "NoDec") denotes that the decoherence correction is used (or not).

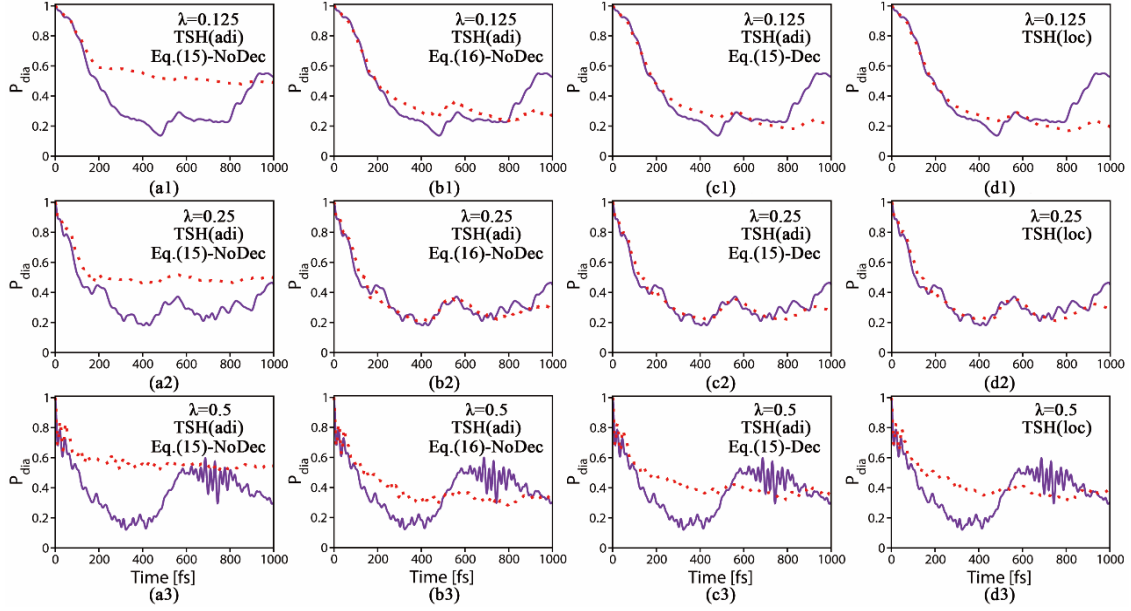

**Figure 5.** Time-dependent diabatic population dynamics at different couplings: the diabatic electronic population of the MCTDH result (violet full line) and the TSH result (red dotted line). Starting from Model III, we rescale the SOC coupling and the scaling factor λ is given in each subfigure. The label "adi" denotes the TSH-adi approach. The label "loc" denotes the TSH-loc approach. The label "Dec" (or "NoDec") denotes that the decoherence correction is used (or not). Here the diabatic population is computed via Eq. (15) or Eq. (16), see the relevant labels in each subfigure.

## C. Initial Nuclear Position

Next, we attempt to shift the initial nuclear coordinates and examine the relevant TSH dynamics in Model III. In the previous section, the initial nuclear sampling is performed according to the ground vibrational level of the ground electronic state. Starting from the initial condition, we simply shift the initial nuclear coordinate along the $v_{30}$ mode by the displacement values $\Delta = 2, 3$ and $4$, and the diabatic potential energy curves along $Q_{30}$ are shown in Appendix IV. We take $Q_{30}$ here because this mode displays the fastest motion in the current Hamiltonian. This shift creates the larger kinetic energy or velocity when the system approaches the *S-T* crossing area of Model III. The time-dependent adiabatic population of the initially populated state is shown in Figure 6. Herein we only show some typical situations, and the other situations are collected in Appendix V.

As discussed in the previous section, the completed dynamics are essentially adiabatic in the model with strong *S-T* coupling, when the dynamics starts from the initial condition near the equilibrium geometry. When the displacement along $Q_{30}$ is made, the nonadiabatic transition starts to become important. With more displacement along $Q_{30}$, the system obtains higher kinetic energy when accessing the *S-T* crossing. The entire nonadiabatic coupling term, in principle, the dot product of the nonadiabatic coupling and velocity, should become larger, thus resulting in the more significant nonadiabatic transition. In this higher velocity situation, the TSH-adi dynamics seem to provide a very good description of the short-time dynamics, although the later population recurrence in the MCTDH result seems not to be reproduced by the TSH-adi dynamics. The current observations on the good performance of the TSH-adi approach in the high kinetic energy cases is very consistent with previous findings that the TSH-adi dynamics generally perform well in the high collision energy situation with the scattering models[28, 119]. However, this observation is only a qualitative conclusion. If the Stuckelberg oscillation exists[8, 41, 42], this conclusion is no longer valid. When the system size increases, the density of the state may become rather high. In this case, the Stuckelberg oscillation may not readily appear. Thus, we believe that the current finding may be more valid in the high-dimensional system.

In the current model, we expect that the high kinetic energy certainly suppresses the possibility of the frustrated hops. This may also be a reason for the better performance of the TSH dynamics.

In Figure 6, the decoherence correction seems not to strongly influence the final results in these situations. When the TSH-loc approach is taken, we receive essentially the same results as the ones obtained by the TSH-adi calculations.

When the diabatic population is examined, conclusions similar to the above findings can be drawn; see Appendix V.

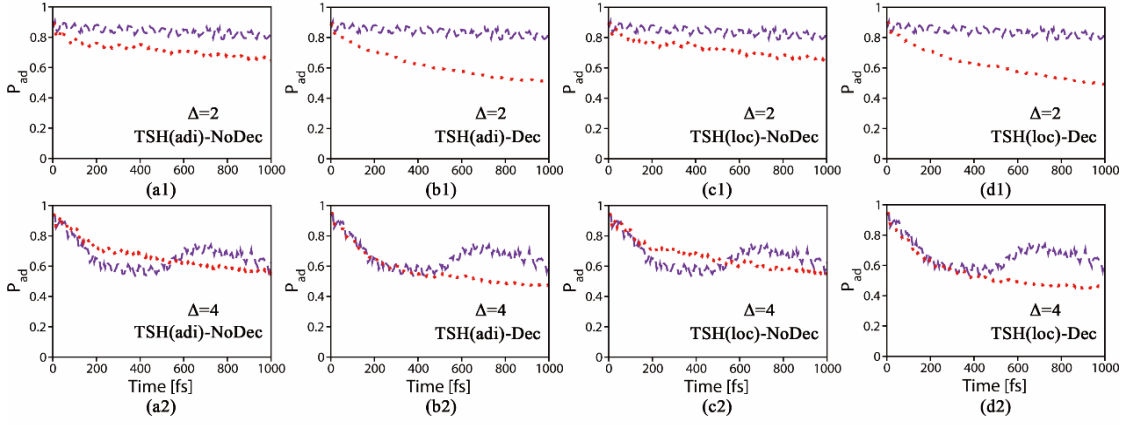

**Figure 6.** Time-dependent adiabatic population dynamics starting from different initial coordinates (2 and 4): the adiabatic electronic population of the MCTDH result (violet dashed line) and the state occupation of the TSH result (red dotted line). Starting from Model III, we shift the initial coordinate along $Q_{30}$ and the shifting value $\Delta$ is given in each subfigure. The label "adi" denotes the TSH-adi approach. The label "loc" denotes the TSH-loc approach. The label "Dec" (or "NoDec") denotes that the decoherence correction is used (or not). Here, we only show two cases, and the rest can be found in Appendix V.

Until now, the TSH dynamic (TSH-adi and TSH-loc) methods have apparently performed quite well in the description of the nonadiabatic ISC dynamics with weak diabatic SOC (increased local crossing in the adiabatic representation) and high kinetic energy conditions. Thus, in principle, if these two conditions are combined together, we should expect that the TSH method should produce effective results comparable to those of the MCTDH approach. This idea is confirmed by further TSH dynamics (TSH-adi and TSH-loc) simulations (see Figure 7), in which we adjust the SOC by the factor of $\lambda = 0.125$ in Model III and shift the initial coordinates along $Q_{30}$ by the value of $\Delta = 4$.

Figure 7 demonstrates that the rapid oscillations of the time-dependent adiabatic population predicted by the MCTDH calculations are effectively reproduced by the TSH dynamics no matter whether the TSH-adi or TSH-loc calculations are performed. After the inclusion of decoherence corrections, the improvement of the TSH (TSH-adi and TSH-loc) performance can be observed.

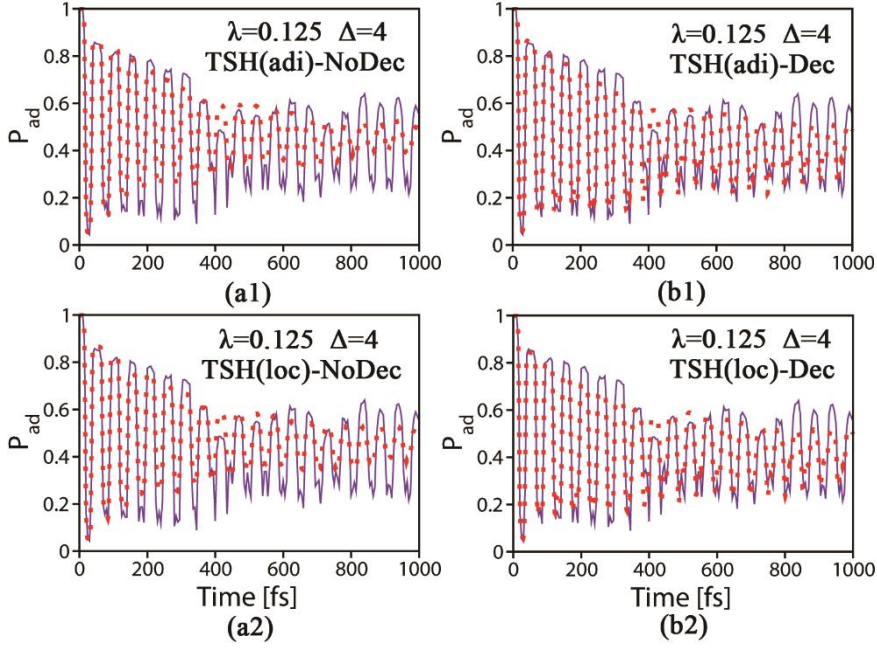

**Figure 7.** Time-dependent adiabatic population dynamics under mixing conditions: the label "adi" denotes the TSH-adi approach, the label "loc" denotes the TSH-loc approach, and the label "Dec" (or "NoDec") denotes that the decoherence correction is used (or not).

### D. Revisiting the TSH Method in Diabatic Representation

In Secction III(A), we show that the TSH-dia calculations do not produce reasonable results in Models I-IV. However, these models with strong SOC represent the situation in which the system dynamics behaves in a rather "adiabatic" fashion due to the large energy gap. In this case, even the TSH-adi approach does not work well. However, the TSH-adi approach works well in the cases with weak SOC and higher kinetic energy. Thus, it is necessary to check whether these cases (suitable for TSH-adi) can be adequately described by the TSH-dia calculations or not.

We considered two models for such a test. The first model include small SOC (by rescaling the SOC by the scaling factor $\lambda = 0.125$ on the basis of Model III). In the second model, the above rescaled SOC was taken and the initial coordinates of all sampled points were shifted along $Q_{30}$ by the value of $\Delta = 4$. Figure 8 shows the ISC dynamics in these models.

Next, we performed TSH-adi and TSH-dia calculations for comparison. In the TSH-adi calculations, we employed a more accurate approach (via Eq. (16)) to calculate the diabatic population. As expected, the TSH-adi dynamic leads to very good results for these two models, see Figure 8(a) and (b). Interestingly, the TSH-dia dynamics produce qualitatively acceptable results in these models. Overall, the TSH-adi method should be recommended.

We also noticed that the performance of TSH-dia dynamics becomes significantly better when high initial energy is employed; see Figure 8(a) and (b). For example, when the large initial shift of the $Q_{30}$ exists, the result becomes comparable with those obtained by the TSH-adi approach and even the full quantum dynamics. This feature was also noticed by Tully in his previous treatment of the scattering problem[28].

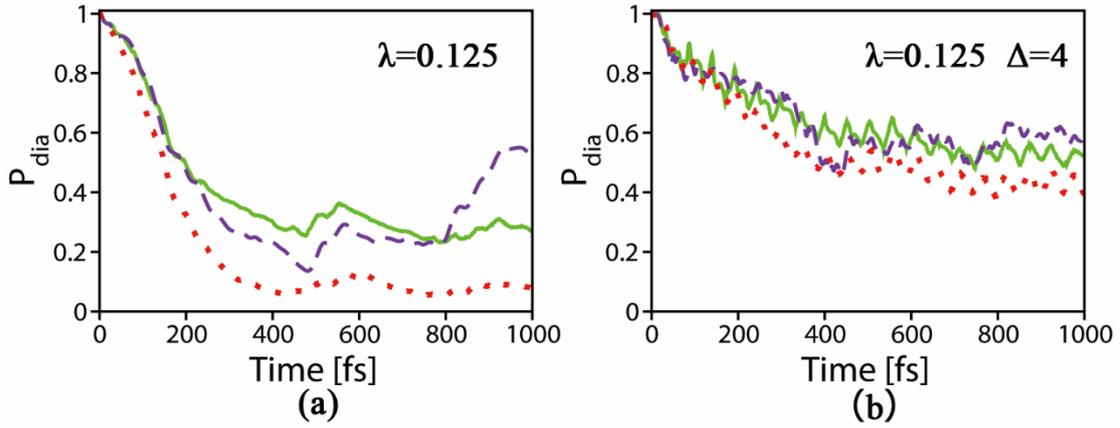

**Figure 8.** Time-dependent diabatic population dynamics in the diabatic representation for some special cases: the diabatic electronic population of the MCTDH result (violet dashed line), the TSH-dia result (red dotted line) and the TSH-adi result (via Eq. (16)) (green full line). Starting from Model III, we rescale the SOC and shift the initial coordinate along $Q_{30}$. The relevant scaling factor λ and shifting value Δ are provided with in each subfigure.

## IV. Conclusion

In this work, we wished to study the performance of the fewest switches TSH method in the ultrafast intersystem crossing processes. The current work focuses on different implementation details and TSH-relevant derivative approaches, such as TSH in the adiabatic or diabatic representations, the local diabatization approach, etc. We took the diabatic model derived from [Re(Br)(CO)$_3$bpy] and employed some simplifications for this purpose. After that, the four-state four-mode Hamiltonian was built to check the performance of the TSH method by comparison with the exact MCTDH dynamics. Here, is a list of our observations.

- The fewest switches TSH dynamics in the spin-adiabatic representation (TSH-adi) and by the local diabatization approaches (TSH-loc) produce very similar results.

- The decoherence corrections in both TSH-adi and TSH-loc dynamics certainly improve the internal consistency between the state population and occupation. If we focus on the reliability of the state occupation, whether the decoherence correction gives better results or not is dependent on the system under study. However, in all of our employed models, the inclusion of the decoherence correction seems not to strongly change the final overall results. Because the internal consistency is reached, it is always acceptable to add such corrections into the treatment of the ISC dynamics.

- In the TSH-adi approach, there are two ways to construct the diabatic population. As recommended by previous works[40, 62], it is suitable to employ Eq. (16) to compute the diabatic population. This offers a useful idea on how to recover the electronic population in the spin-diabatic basis (such as the population of the singlet or triplet states).

- In the strong SOC models (large diabatic coupling), the very large adiabatic energy gap appears and the dynamics become more "adiabatic" in nature. The TSH-adi and TSH-loc may overestimate the decay rate from the upper adiabatic state to the lower adiabatic state.

- With the weak diabatic coupling, the models yield a small adiabatic energy gap and very local nonadiabatic coupling. In this case, both the TSH-adi and TSH-loc approaches produce rather satisfactory results. Thus, it is reasonable to employ them to study the ultrafast ISC dynamics with weak spin-orbital couplings. However, we also need to use caution here because if the SOC becomes extremely small or completely zero, the study of the relevant ISC dynamics becomes meaningless.

- When the trajectory passes the crossing region with high velocity, the TSH-adi and TSH-loc dynamics work well. This means that the ISC dynamics in the chemical reactions under the high initial energy may be captured by the TSH simulation. Certainly, this is only a very rough and qualitative finding. In the presence of the Stuckelberg oscillations, it is well known that the situation should be more complicated.

- The performance of the TSH-dia approach is generally worse than the performance of the TSH-adi and TSH-loc approaches. In the large diabatic coupling case, the TSH-dia method may produce results which are far from the correct ones. In the case with very small diabatic coupling and high kinetic energy, the TSH-dia dynamics may provide qualitative reasonable results for the population dynamics, for instance showing the decay time scale with the same magnitude as the current one. Thus, the TSH-adi and TSH-loc approaches are always recommended. Only in the models with very weak SOC and very high kinetic energy, the TSH-dia may represent a secondary choice for qualitative understanding of the ISC processes.

Although some of the above points were mentioned in previous works, we notice that these available works focus on different models, such as conical intersections, one-dimensional scattering problems and a model composed of harmonic potentials in both the heavier and lighter particle coordinates[28, 39, 69, 73-75, 115, 116]. At the same time, many of these works primarily focus on the final reaction rates. Instead, our work is more concerned with the time-dependent feature of the nonadiabatic ISC dynamics based on a linear vibronic S-T coupling model. This provides us with more direct understanding of the TSH dynamics in the treatment of the ultrafast ISC dynamics and other similar ultrafast dynamics. In the current work, we explicitly point out the above findings, and we believe that they are very useful to guide the future study of the ISC dynamics with the TSH method. Ultimately, we fully understand that the TSH dynamics may not solve all problems at once, thus it is also very important to develop additional mixed quantum-classical or semiclassical dynamics approaches for the reliable and efficient description of the realistic nonadiabatic intersystem crossing dynamics.

**Acknowledgments**

This work is supported by NSFC Projects (No. 21673266, 21873112 and 21503248). The authors thank the Supercomputing Center of the Computer Network Information Center, CAS; the National Supercomputing Center in Shenzhen; the National Supercomputing Center in Guangzhou and the Supercomputing Center of CAS-QIBEBT for providing computational resources.

**Appendix**

I.  Additional results of the TSH-dia and TSH-adi approaches

The nonadiabatic ISC dynamics in Model III are shown in Figure 9.

In the TSH-dia approach, two triplet states retain the same populations (Figure 9. (a)), because their couplings with the singlet state are conjugated to each other. The last triplet state does not exhibit any population because it does not couple with the singlet state.

In the TSH-adi approach, only two adiabatic states are involved, while another two states do not exhibit any population throughout the entire dynamics (Figure 9. (b)).

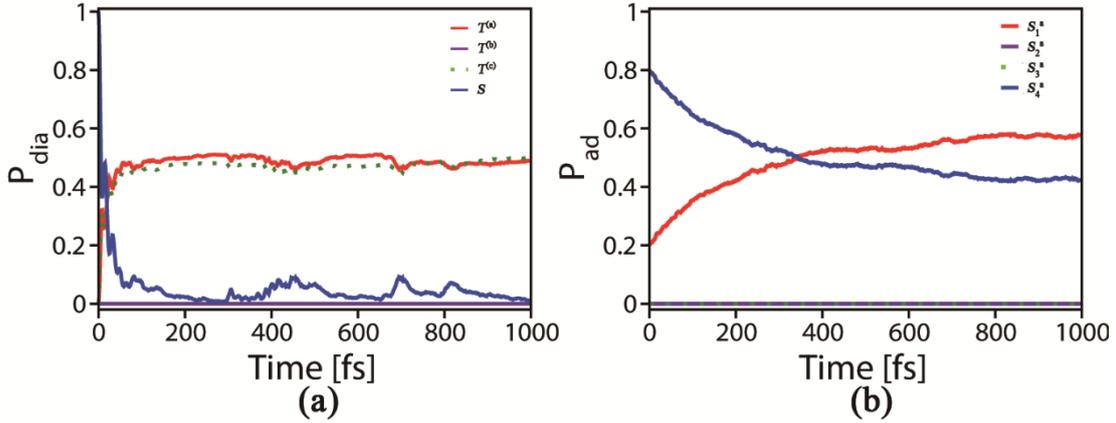

**Figure 9.** The nonadiabatic ISC dynamics for Model III including all electronic states: (a) the diabatic state occupation obtained from the TSH-dia dynamics, (b) the adiabatic state occupation obtained from the TSH-adi dynamics.

II. Frustrated hops in the TSH dynamics in Model III

When the TSH-adi dynamics calculations are performed in the adiabatic representation for Model III, each trajectory experiences many hopping events, while not all of the hops are allowed due to the existence of frustrated hops. The distribution of the number of frustrated hops for each trajectory is given in Figure 10. When the decoherence correction is not included, there is a large number of frustrated hops. However, the inclusion of the decoherence correction largely reduces the frustrated hops.

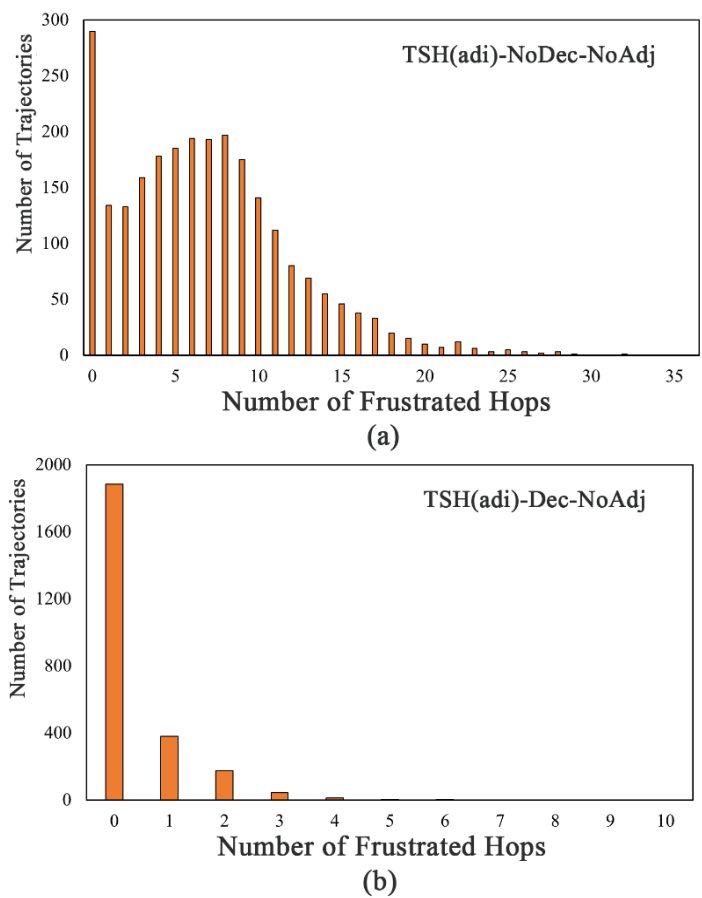

**Figure 10.** The number of frustrated hops in the TSH-adi dynamics of Model III: (a) without the decoherence correction, (b) with the decoherence correction. The velocity remains unchanged when the frustrated hop happens.

III. "Typical" trajectory of Model III

In the TSH-adi simulation of Model III, a "typical" trajectory is given in Figure 11. It is clear that the system experiences the fast Rabi-type oscillation. Because the electronic population change quickly, the sudden hop may happen very easily.

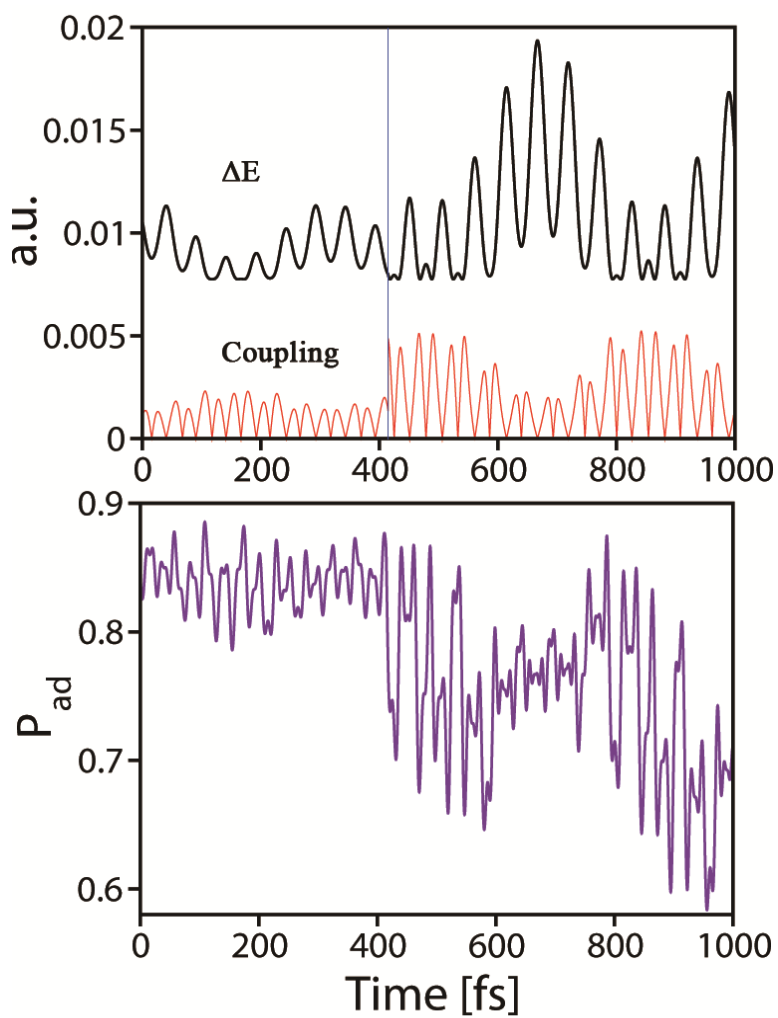

**Figure 11.** Behavior of a "typical" trajectory of Model III: Bottom panel: adiabatic population. Top panel: energy difference (black line) and the module of dynamic couplings (red line). The vertical bar indicates a surface hopping event.

IV. The diabatic potential energy lines along $Q_{30}$ for Model III

We considered different initial conditions and put the initial sampling geometry distribution centered at different values of $Q_{30}$. To get a rather clear idea on where these initial conditions are located, we plot the diabatic potential energy surface along $Q_{30}$ for Model III, see Figure 12.

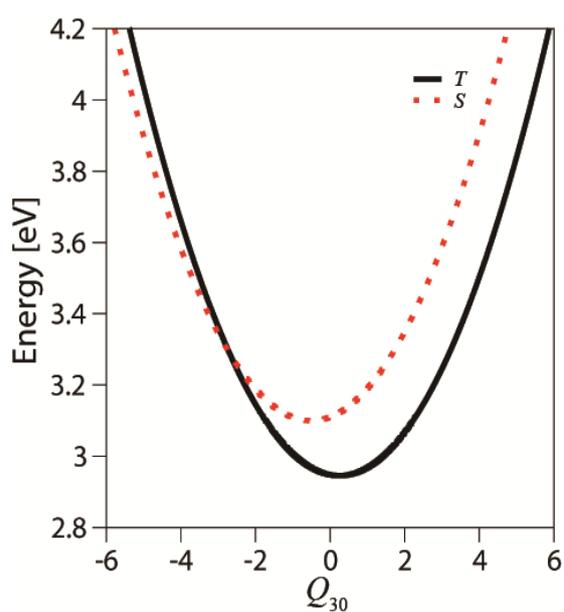

**Figure 12.** The diabatic potential energy lines along $Q_{30}$ for Model III.

## V. Nonadiabatic ISC dynamics with different initial nuclear coordinates

As discussed in the main text, the TSH dynamics tend to perform very well when the trajectory accesses the *S-T* crossing with high velocities, due to the larger shift of the initial coordinate along $Q_{30}$; see Figures 13 and 14.

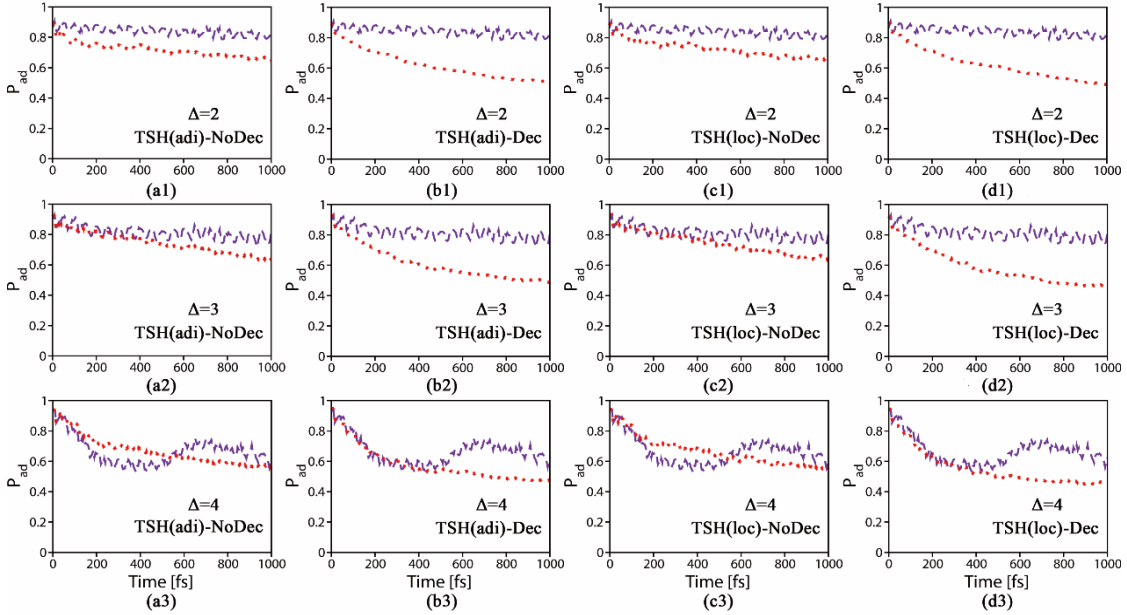

**Figure 13.** Time-dependent adiabatic population dynamics starting from different initial coordinates: the adiabatic electronic population of the MCTDH result (violet dashed line) and the state occupation of the TSH result (red dotted line). Starting from Model III, we shift the initial coordinate along $Q_{30}$, and the shifting value Δ is given in each subfigure. The label "adi" denotes the TSH-adi approach. The label "loc" denotes the TSH-loc approach. The label "Dec" (or "NoDec") denotes that the decoherence correction is used (or not).

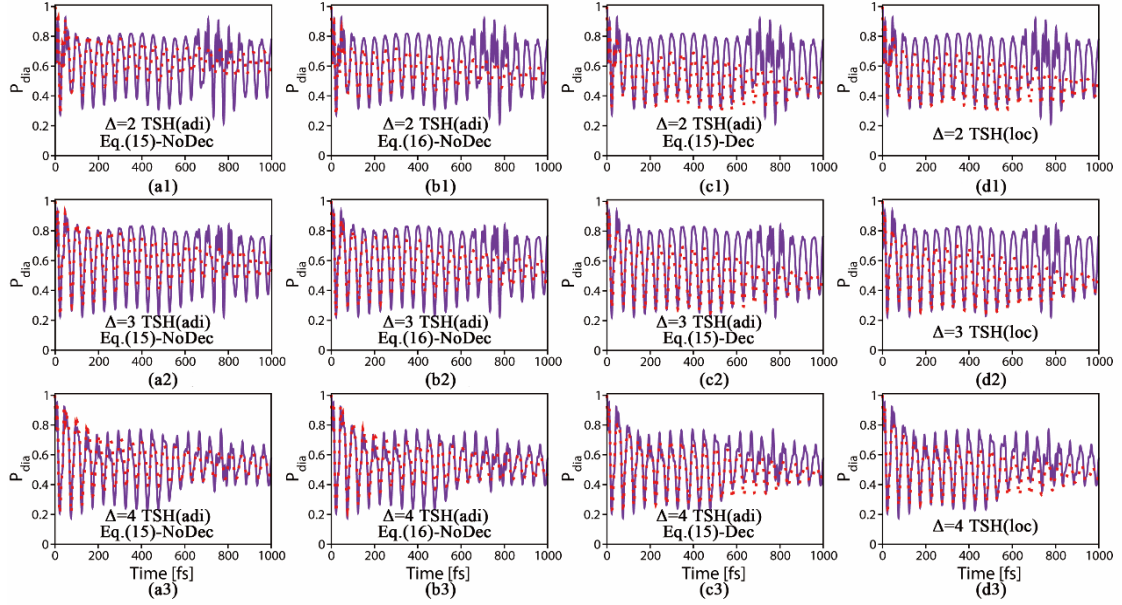

**Figure 14.** Time-dependent diabatic population dynamics starting from different initial coordinates: the diabatic electronic population of the MCTDH result (violet full line) and the TSH result (red dotted line). Starting from Model III, we shift the initial coordinate along $Q_{30}$, and the shifting value $\Delta$ is given in each subfigure. The label "adi" denotes the TSH-adi approach. The label "loc" denotes the TSH-loc approach. The label "Dec" (or "NoDec") denotes that the decoherence correction is used (or not). Here the diabatic population is computed via Eq. (15) or Eq. (16); see the relevant labels in each subfigure.

## VI. Treatment of frustrated hops

Different approaches were proposed to treat frustrated hops. For example, the frustrated hops may be simply neglected, with no further treatment employed[70]. Alternatively, it is also possible to adjust the momentum by considering the similarity between the elastic scattering and the frustrated hop[33]. The momentum is composed of two components, one perpendicular to the nonadiabatic coupling and the other one parallel to it. The parallel component is then adjusted to its reversed direction, while the perpendicular component remains unchanged. Certainly, energy conservation must be satisfied in such momentum adjustments. We also noticed that more advanced approaches were proposed[120-123].

We execute nonadiabatic ISC dynamics calculations with the above two different treatments of frustrated hops, and the relevant results are shown in Figure 15 and Figure 16. It is seen that the TSH-adi dynamics may be modified by different methods to treat frustrated hops. When no decoherence correction is added, we found that the first method of neglecting frustrated hops seems to provide better results for Model III, and a similar finding was obtained by the previous work by Stock and coworkers[70]. Certainly, the proper approach to treat the frustrated hops may be system dependent, because many previous works also suggest that the second way is superior[120-123]. We expect that more discussions on this topic can be found in available references[54, 120-123]. This complicity is also shown in the current simulation, because two different ways seem to predict very similar results for the model with small diabatic coupling.

However, when the decoherence correction is added into the TSH-adi dynamics, two different treatments of the frustrated hops produce very similar results in the current models.

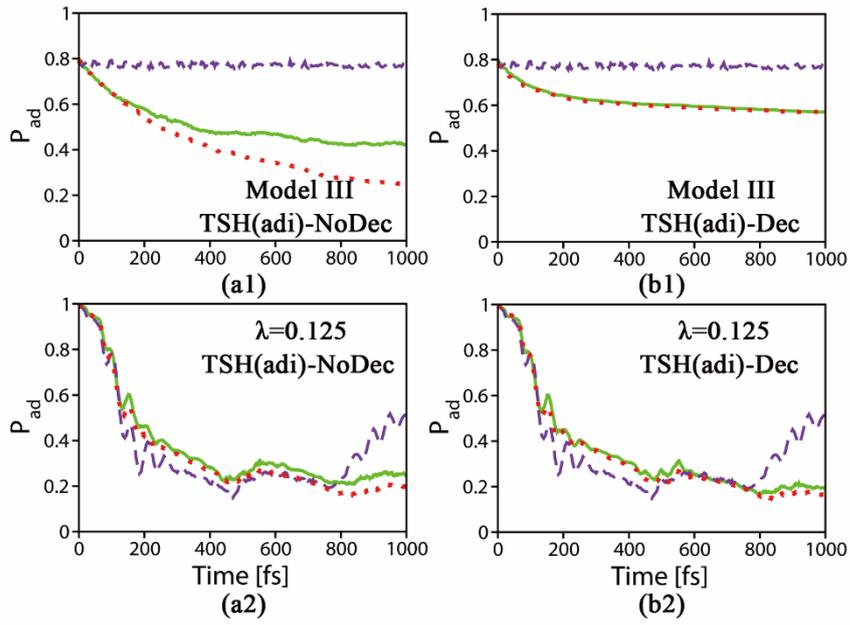

**Figure 15.** Time-dependent population dynamics with different treatments of frustrated hops in Model III and the situation where the scaling factor $\lambda = 0.125$; the adiabatic electronic population of the MCTDH result (violet dashed line) and the adiabatic state occupation of the TSH result with (red dotted line) and without (green full line) modifying the momentum after a rejected hop.

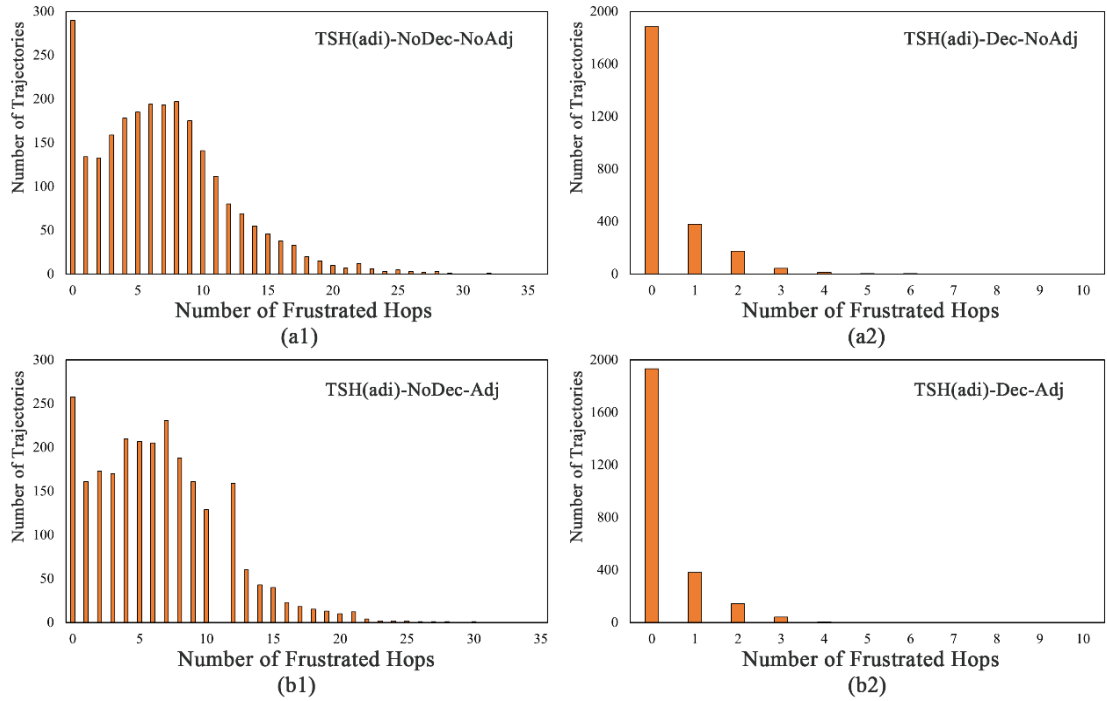

**Figure 16.** The number of frustrated hops in the TSH-adi dynamics of Model III: (a) without changing the velocities and (b) by reversing the component of the velocity along the direction of the nonadiabatic coupling vector for a rejected hop.

## VII. Numerical Stability

In the weak spin-diabatic coupling limit, we need to know whether the current model suffers from the so-called "trivial crossing"[39, 103] problem or not. In the trivial crossing cases, the very small diabatic coupling may result in two adiabatic potential energy surfaces creating the crossing point with a very small energy gap, thus resulting in extremely localized nonadiabatic couplings. In this situation, the trajectory may easily "miss" the crossings and wrongly remain on the same spin-adiabatic state. Thus, it is necessary to verify whether the current sets of the model suffer from the trivial crossing point, when the SOC becomes small. Although several advanced methods to avoid such problems are proposed[39, 103], we simply wish to determine whether the trivial crossing problems happen and to check the numerical stability in our testing models.

We chose Model III with modified parameters ($\lambda = 0.125$, $\Delta = 4$) as a representative and employed the simplest approach by changing the time step for the integration of the nuclear motion or the electronic motion. When different nuclear time steps or electronic time steps were used, the results were not very sensitive to the selection of the time step, see Figure 17 and 18. Thus, the trivial crossing problem does not exist in our testing models here.

When the large time step 0.5 fs is used to propagate the nuclear motion, the minor difference appears in Figure 17 for the TSH-adi approach, while the results are still acceptable. As a contrast, the TSH-loc approach still exhibits the excellent agreement with respect to the smaller time steps.

If we further employ the large step for nuclear propagation, such as 0.8 fs, the total energy is no longer conserved in both TSH-adi and TSH-loc methods.

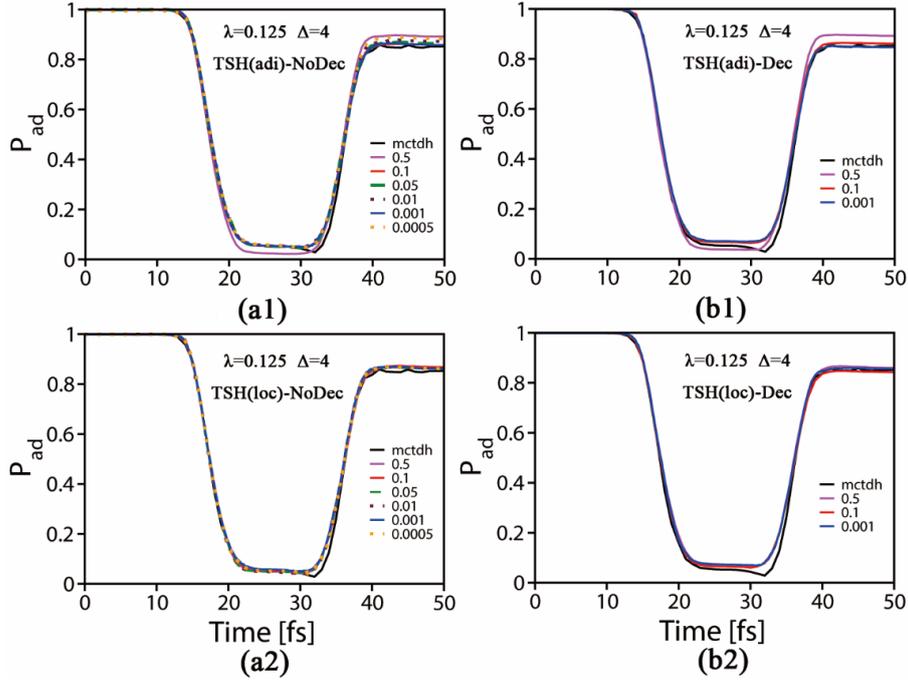

**Figure 17.** Time-dependent adiabatic population dynamics under different nuclear time steps: the label "adi" denotes the TSH-adi approach. The label "loc" denotes the TSH-loc approach. The label "Dec" (or "NoDec") denotes that the decoherence correction is used (or not). The time step (unit fs) for the integration of the nuclear motion is given in each subfigure. The time step for the integration of the electronic motion is 0.001 fs in most situations. The same time step is used for both nuclear and electronic motion when the time step is 0.0005 fs.

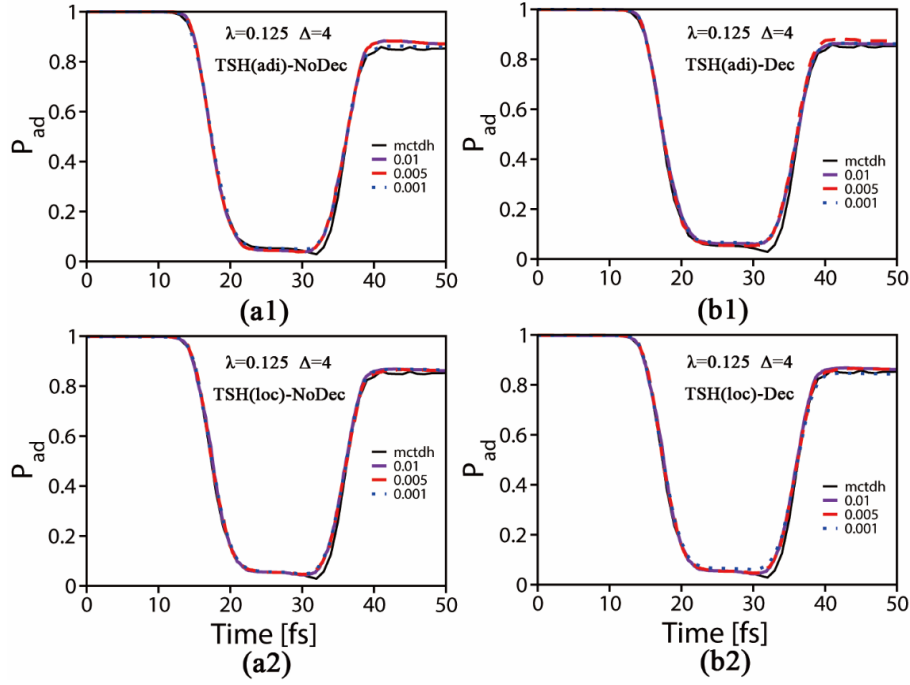

**Figure 18.** Time-dependent adiabatic population dynamics under different electronic time steps: the label "adi" denotes the TSH-adi approach. The label "loc" denotes the TSH-loc approach. The label "Dec" (or "NoDec") denotes that the decoherence correction is used (or not). The time step (unit fs) for the integration of the electronic motion is given in each subfigure. The time step for the integration of the nuclear motion is 0.1 fs.

**Table list**

Table 1. List of the parameters in the Model Hamiltonian

| mode | $\omega$ / eV | $\kappa$ / eV | |
|---|---|---|---|
| | | S | T |
| 7 | 0.0116 | 0.0187 | -0.0161 |
| 11 | 0.0188 | 0.0091 | 0.0002 |
| 13 | 0.0229 | -0.0271 | -0.0261 |
| 30 | 0.0792 | 0.0404 | -0.0196 |

| $\eta$ / eV | |
|---|---|
| $T^{(a)}$-$S$ | -0.0719 - 0.0196$i$ |

Table 2. List of transition energy values for Models I-IV

| State | S | T | | | |
|---|---|---|---|---|---|
| Model | | I | II | III | IV |
| $E_{(0)}$ / eV | 3.1100 | 3.0539 | 3.1599 | 2.9479 | 3.2376 |

**Figure List**

**Figure 1**. Diabatic potential energy surface (PES) along $Q_7$ for Models I-IV. The red dashed line denotes the PES of the singlet $S$ state and the back solid line indicates the PES of the triplet $T$ states.

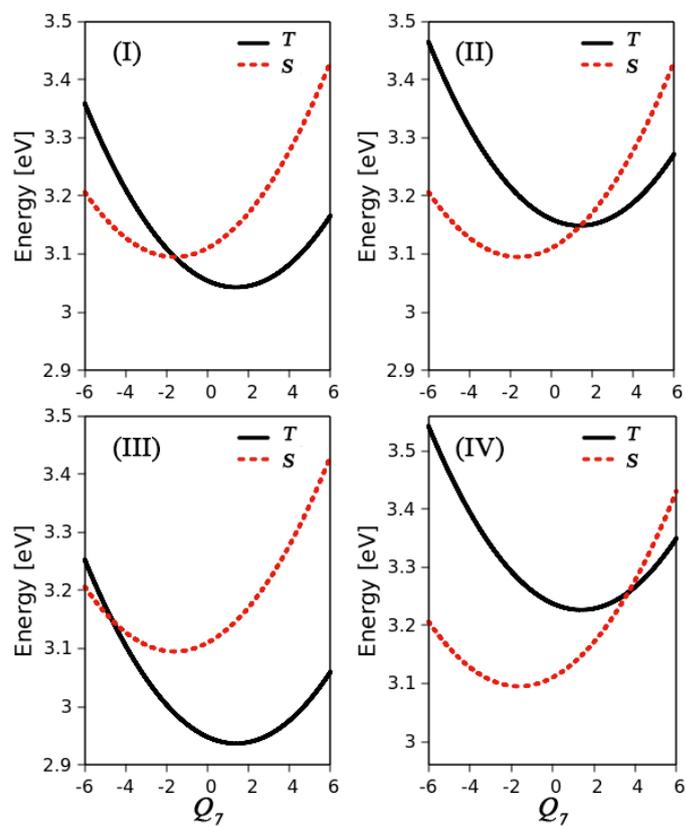

**Figure 2**. Time-dependent diabatic population dynamics: the diabatic electronic population of the MCTDH result (violet dashed line), the diabatic state occupation (red dotted line) and the diabatic electronic population (green full line) of the TSH result.

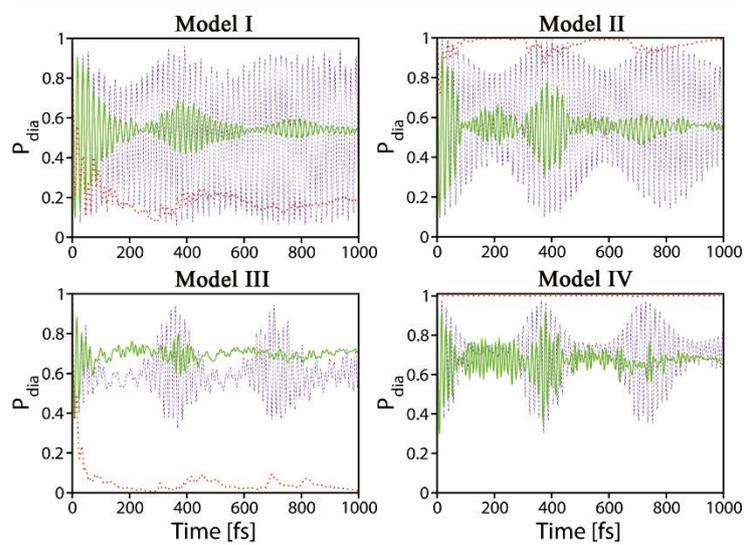

**Figure 3**. Time-dependent adiabatic population dynamics: the adiabatic electronic population of the MCTDH result (violet dashed line), the adiabatic state occupation (red dotted line) and the adiabatic electronic population (green full line) of the TSH results. The models employed are given in each figure. The label "adi" denotes the TSH-adi approach. The label "loc" denotes the TSH-loc approach. The label "Dec" (or "NoDec") denotes that the decoherence correction is used (or not).

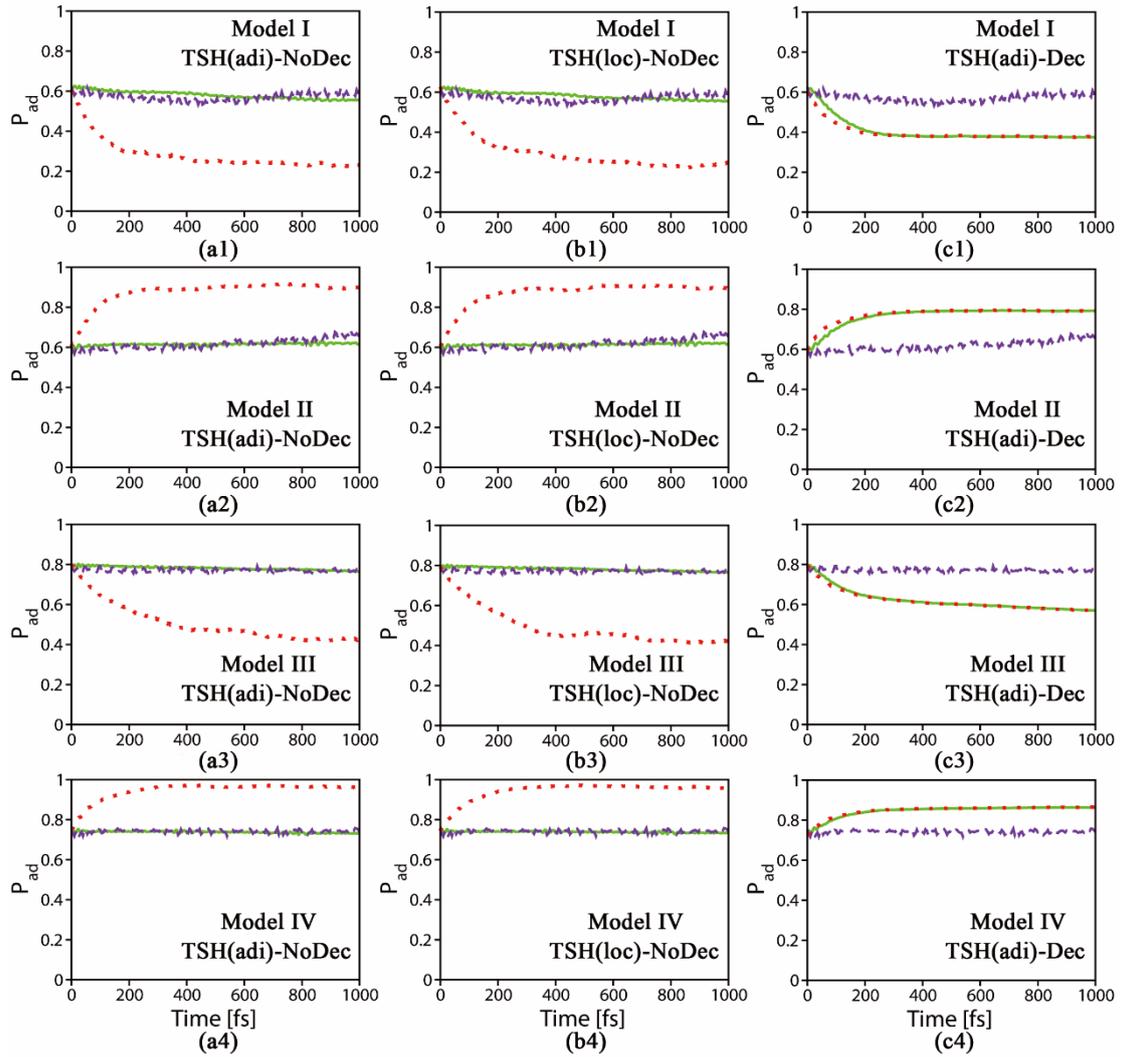

**Figure 4.** Time-dependent adiabatic population dynamics at different couplings: the adiabatic electronic population of the MCTDH result (violet dashed line), the state occupation (red dotted line) and the electronic population (green full line) of the TSH result. Starting from Model III, we rescale the SOC; the scaling factor λ is given in each subfigure. The label "adi" denotes the TSH-adi approach. The label "loc" denotes the TSH-loc approach. The label "Dec" (or "NoDec") denotes that the decoherence correction is used (or not).

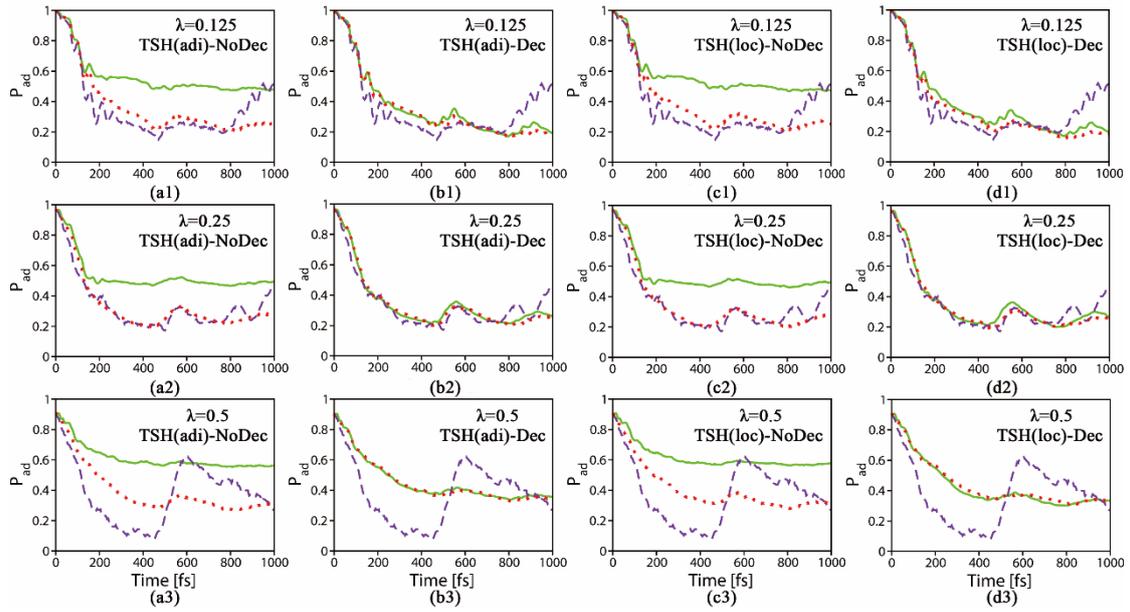

**Figure 5**. Time-dependent diabatic population dynamics at different couplings: the diabatic electronic population of the MCTDH result (violet full line) and the TSH result (red dotted line). Starting from Model III, we rescale the SOC and the scaling factor λ is given in each subfigure. The label "adi" denotes the TSH-adi approach. The label "loc" denotes the TSH-loc approach. The label "Dec" (or "NoDec") denotes that the decoherence correction is used (or not). Here, the diabatic population is computed via Eq. (15) or Eq. (16); see the relevant labels in each subfigure.

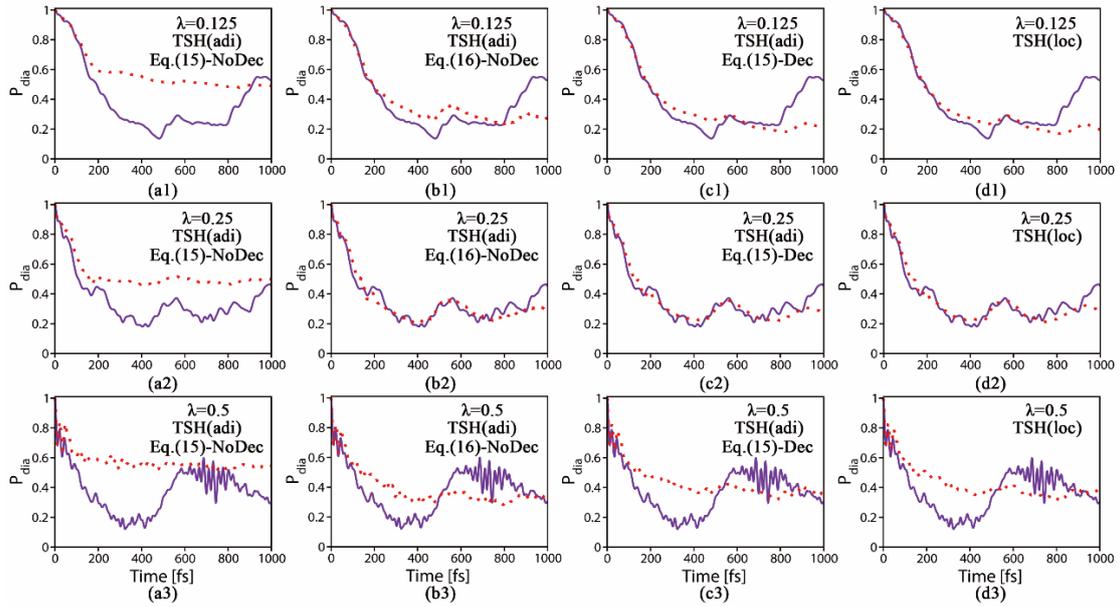

**Figure 6**. Time-dependent population dynamics starting from different initial coordinates (2 and 4): the adiabatic electronic population of the MCTDH result (violet dashed line) and the state occupation of the TSH result (red dotted line). Starting from Model III, we shift the initial coordinate along $Q_{30}$ and the shifting value $\Delta$ is given in each subfigure. The label "adi" denotes the TSH-adi approach. The label "loc" denotes the TSH-loc approach. The label "Dec" (or "NoDec") denotes that the decoherence correction is used (or not).

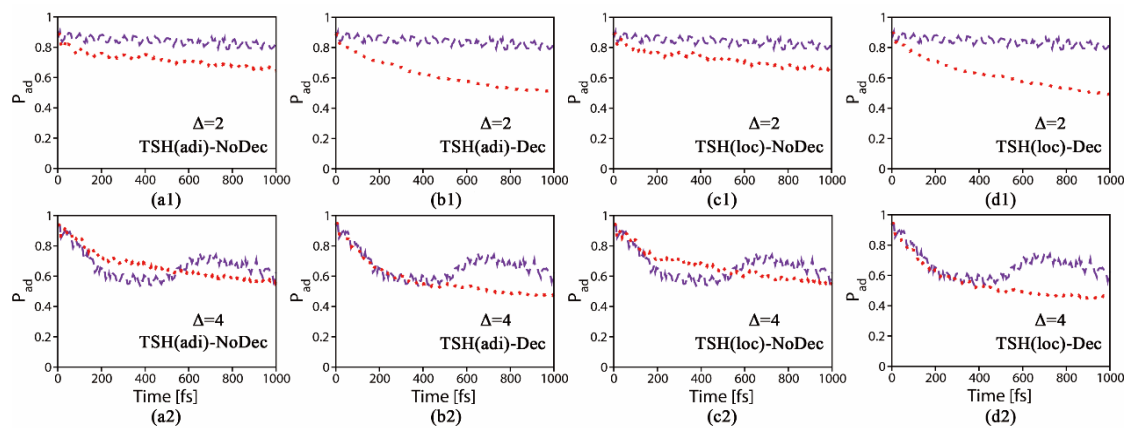

**Figure 7**. Time-dependent adiabatic population dynamics under mixing conditions: the label "adi" denotes the TSH-adi approach, the label "loc" denotes the TSH-loc approach, and the label "Dec" (or "NoDec") denotes that the decoherence correction is used (or not).

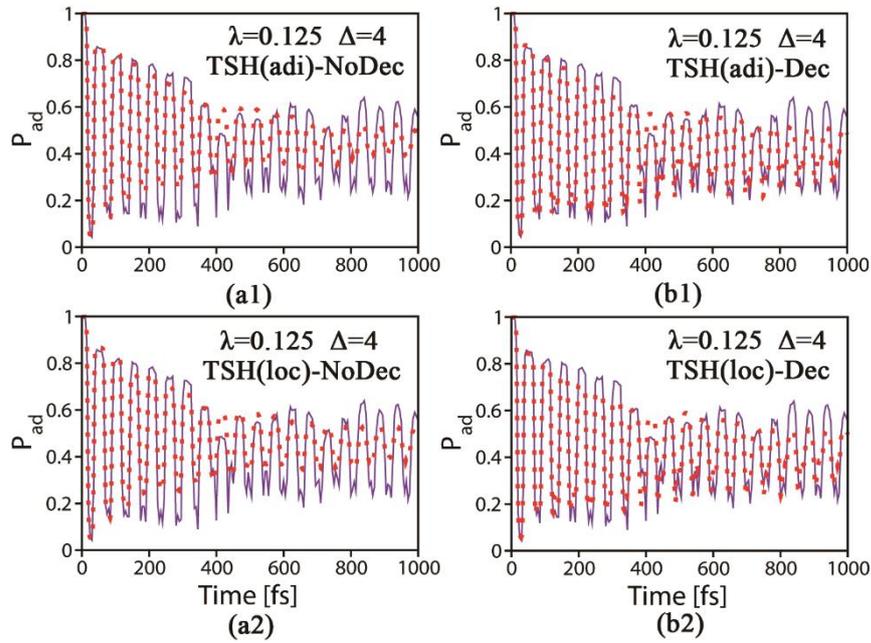

**Figure 8**. Time-dependent diabatic population dynamics in the diabatic representation for some special cases: the diabatic electronic population of the MCTDH result (violet dashed line), the TSH result performed in diabatic representation (red dotted line) and the TSH result performed in the adiabatic representation computing via Eq. (16) (green full line). Starting from Model III, we rescale the SOC and shift the initial coordinate along $Q_{30}$. The relevant scaling factor λ and shifting value Δ are provided with in each subfigure.

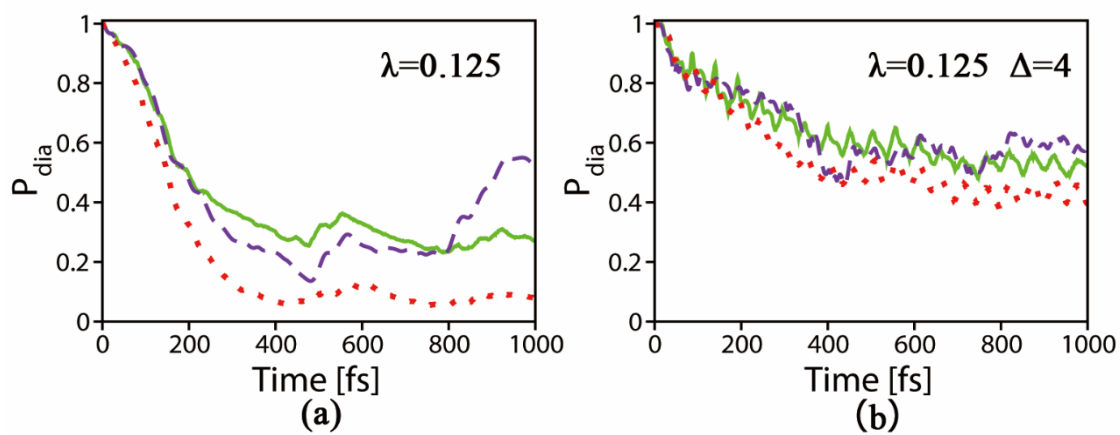

**Figure 9**. The nonadiabatic ISC dynamics for Model III including all electronic states: (a) the diabatic state occupation obtained from the TSH dynamics performed in the diabatic representation, (b) the adiabatic state occupation obtained from the TSH-adi dynamics.

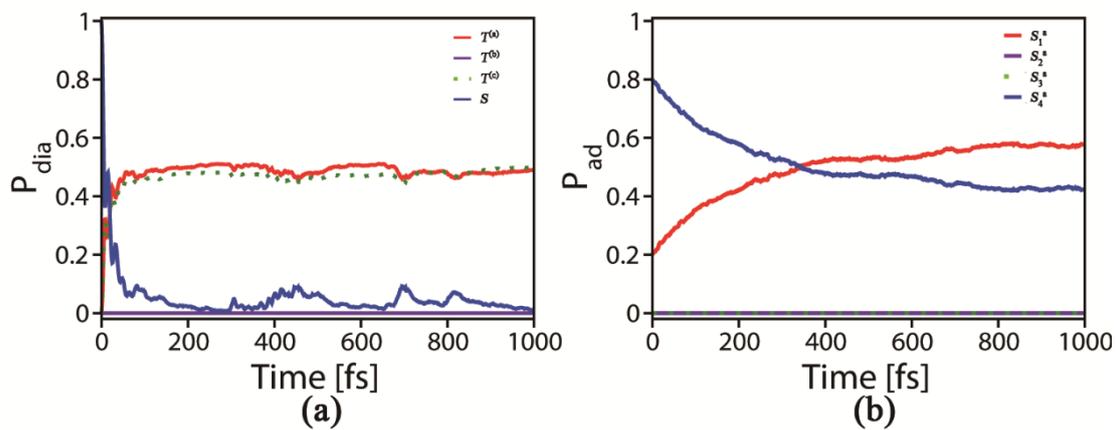

**Figure 10**. The number of frustrated hops in the TSH-adi dynamics of Model III: (a) without the decoherence correction, (b) with the decoherence correction. The velocity remains unchanged when the frustrated hop happens.

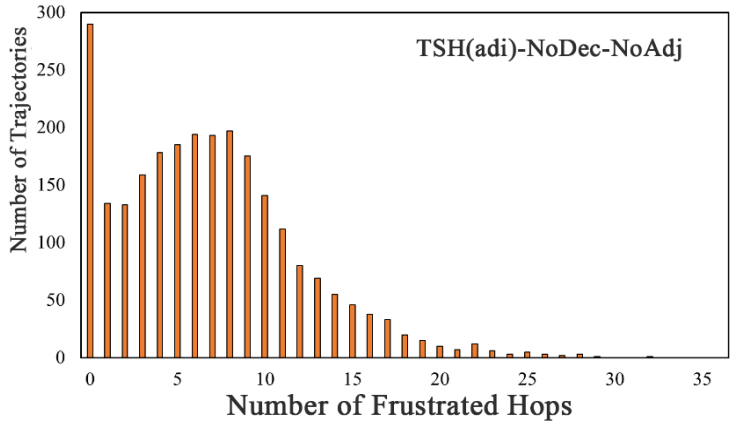

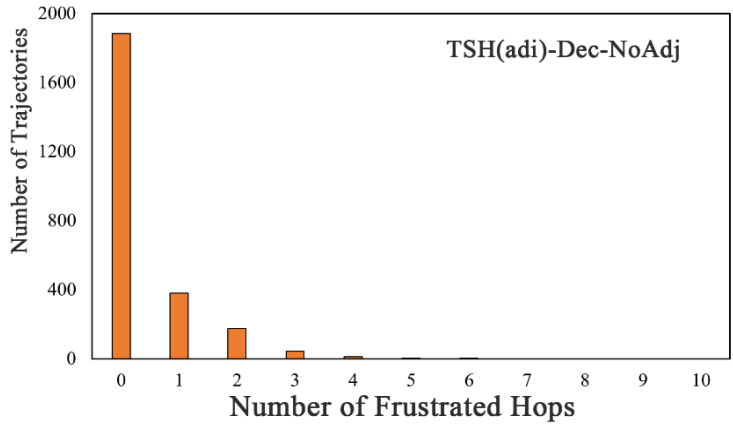

**Figure 11.** Behavior of a "typical" trajectory of Model III: Bottom panel: adiabatic population. Top panel: energy difference (black line) and the module of dynamic couplings (red line). The vertical bar indicates a surface hopping event.

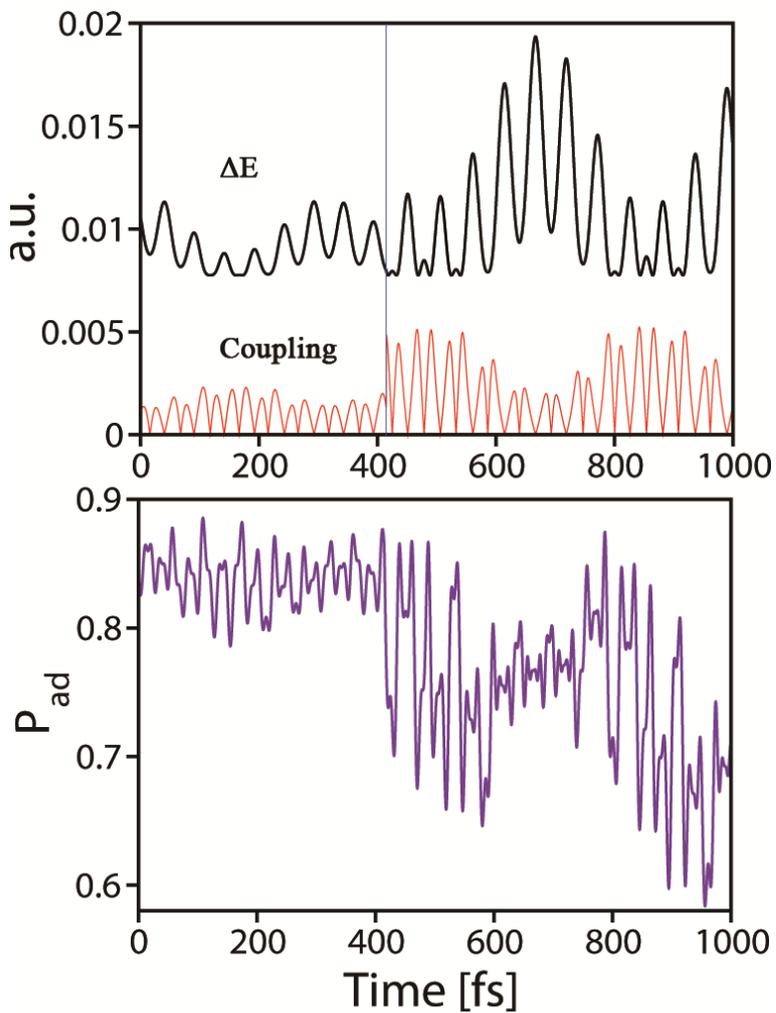

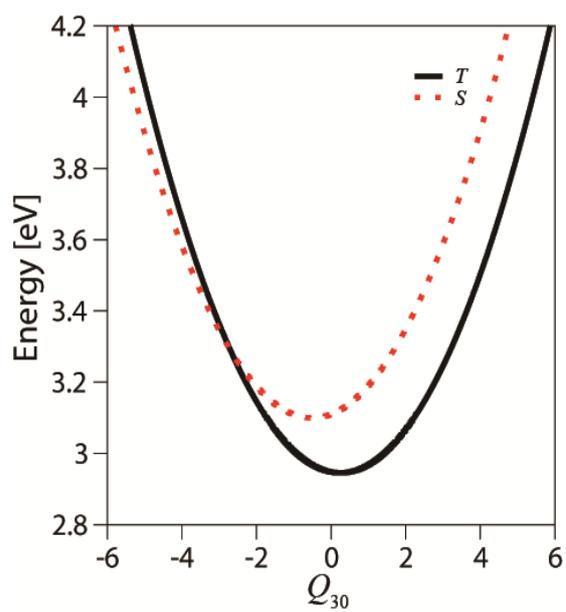

**Figure 12**. The diabatic potential energy lines along $Q_{30}$ for Model III.

**Figure 13**. Time-dependent adiabatic population dynamics starting from different initial coordinates: the adiabatic electronic population of the MCTDH result (violet dashed line) and the state occupation of the TSH result (red dotted line). Starting from Model III, we shift the initial coordinate along $Q_{30}$, and the shifting value $\Delta$ is given in each subfigure. The label "adi" denotes the TSH-adi approach. The label "loc" denotes the TSH-loc approach. The label "Dec" (or "NoDec") denotes that the decoherence correction is used (or not).

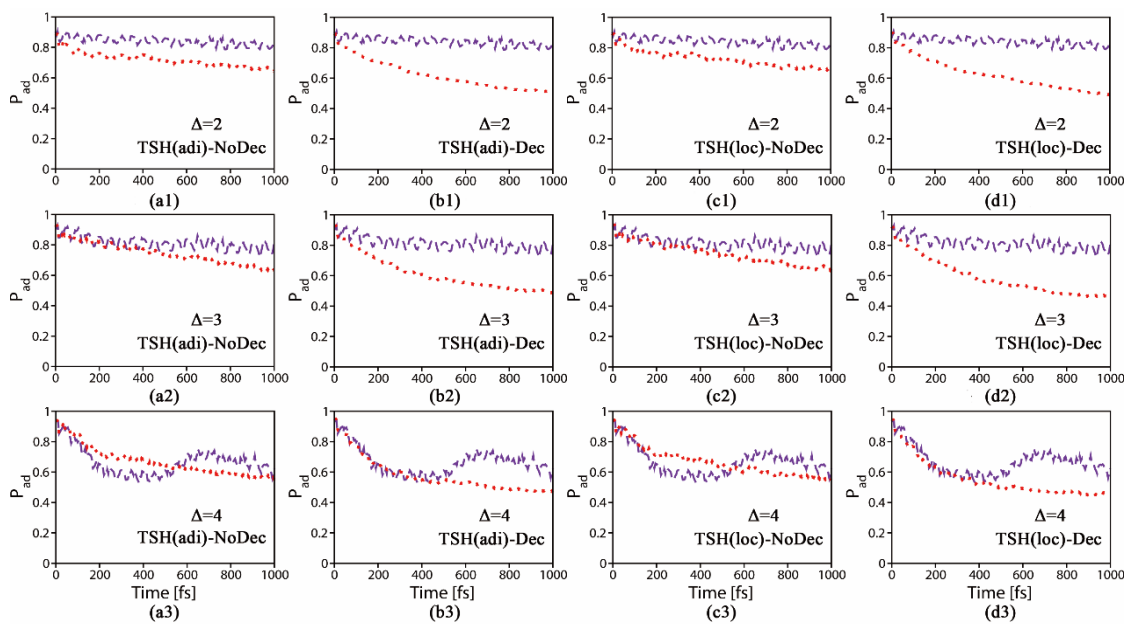

**Figure 14**. Time-dependent diabatic population dynamics starting from different initial coordinates: the diabatic electronic population of the MCTDH result (violet full line) and the TSH result (red dotted line). Starting from Model III, we shift the initial coordinate along $Q_{30}$, and the shifting value $\Delta$ is given in each subfigure. The label "adi" denotes the TSH-adi approach. The label "loc" denotes the TSH-loc approach. The label "Dec" (or "NoDec") denotes that the decoherence correction is used (or not). Here the diabatic population is computed via Eq. (15) or Eq. (16); see the relevant labels in each subfigure.

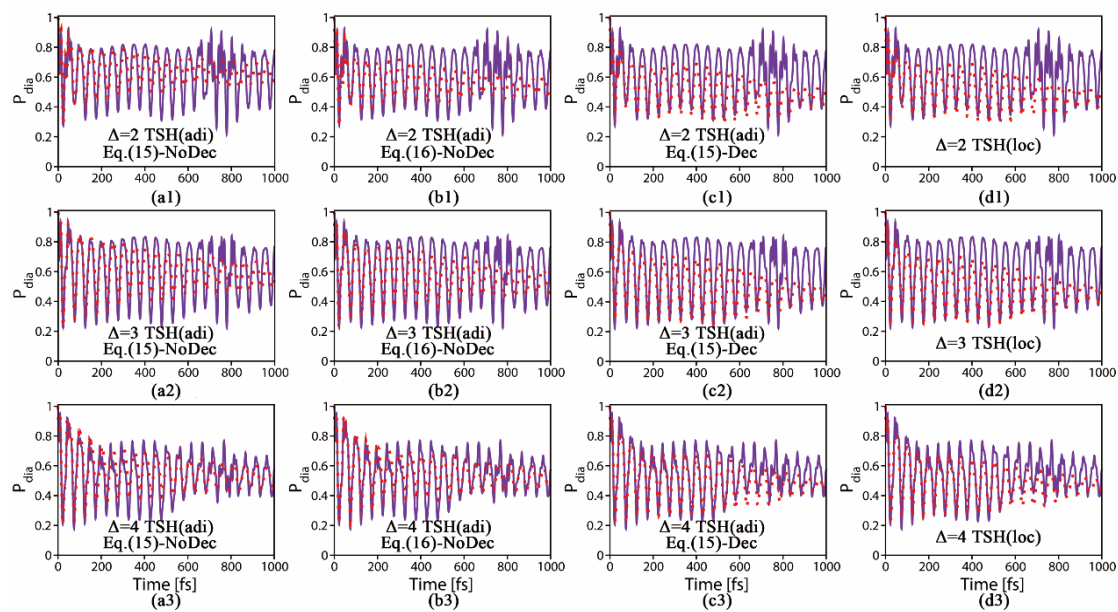

**Figure 15**. Time-dependent population dynamics with different treatments of frustrated hops: in Model III and the situation where the scaling factor $\lambda = 0.125$: the adiabatic electronic population of the MCTDH result (violet dashed line) and the adiabatic state occupation of the TSH result with (red dotted line) and without (green full line) modifying the velocities after a rejected hop.

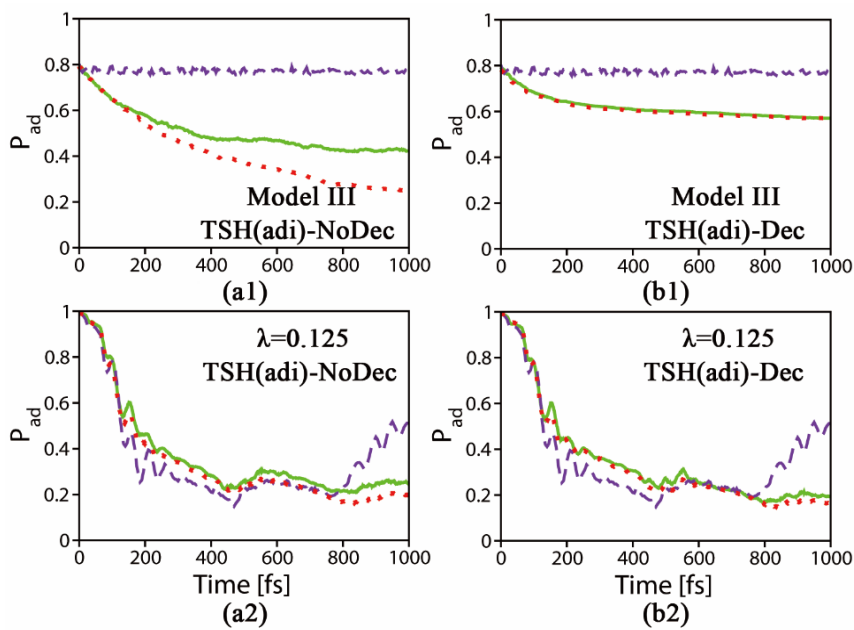

**Figure 16**. The number of frustrated hops in the TSH-adi dynamics of Model III: (a) without changing the velocities and (b) by reversing the component of the velocity along the direction of the nonadiabatic coupling vector for a rejected hop.

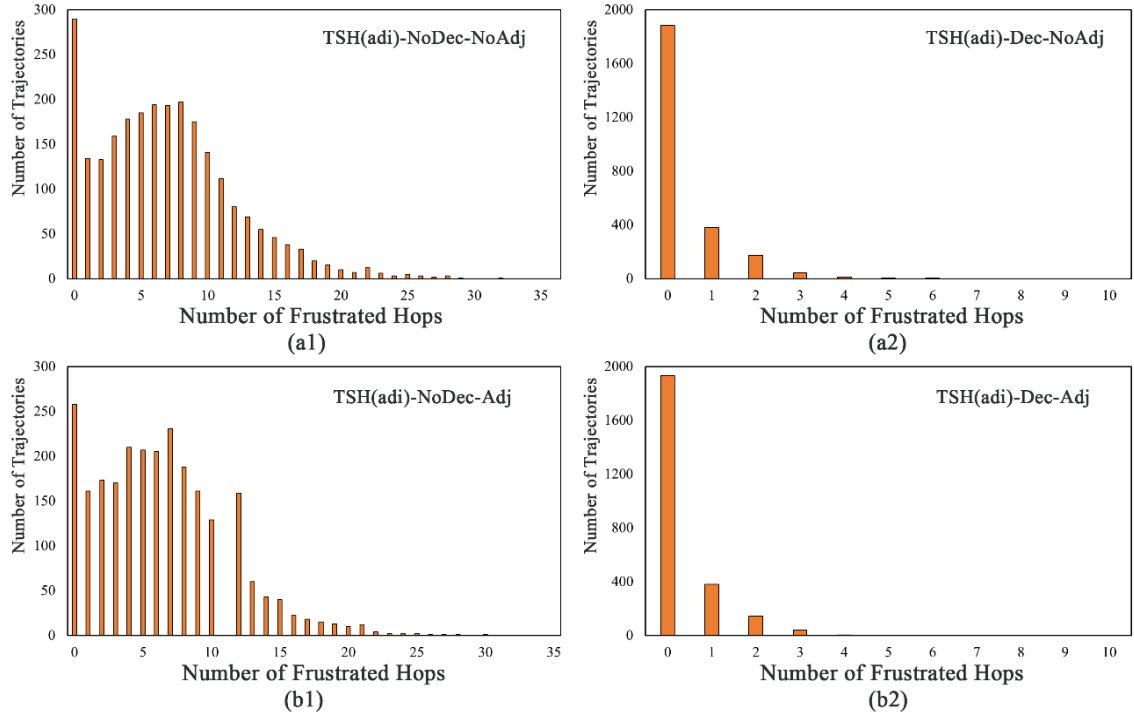

**Figure 17**. Time-dependent adiabatic population dynamics under different nuclear time steps: the label "adi" denotes the TSH-adi approach. The label "loc" denotes the TSH-loc approach. The label "Dec" (or "NoDec") denotes that the decoherence correction is used (or not). The time step (unit fs) for the integration of the nuclear motion is given in each subfigure.

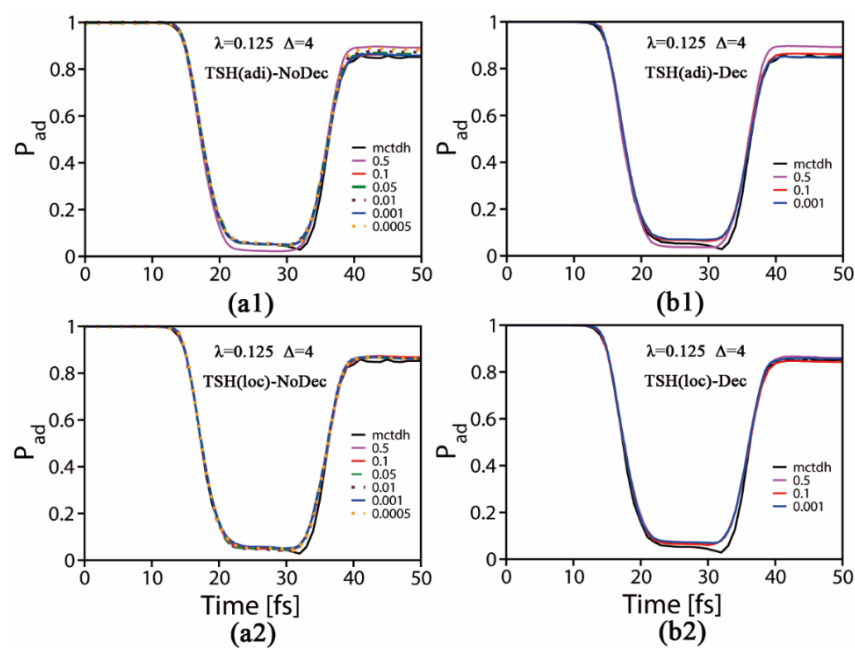

**Figure 18**. Time-dependent adiabatic population dynamics under different electronic time steps: the label "adi" denotes the TSH-adi approach. The label "loc" denotes the TSH-loc approach. The label "Dec" (or "NoDec") denotes that the decoherence correction is used (or not). The time step (unit fs) for the integration of the electronic motion is given in each subfigure.

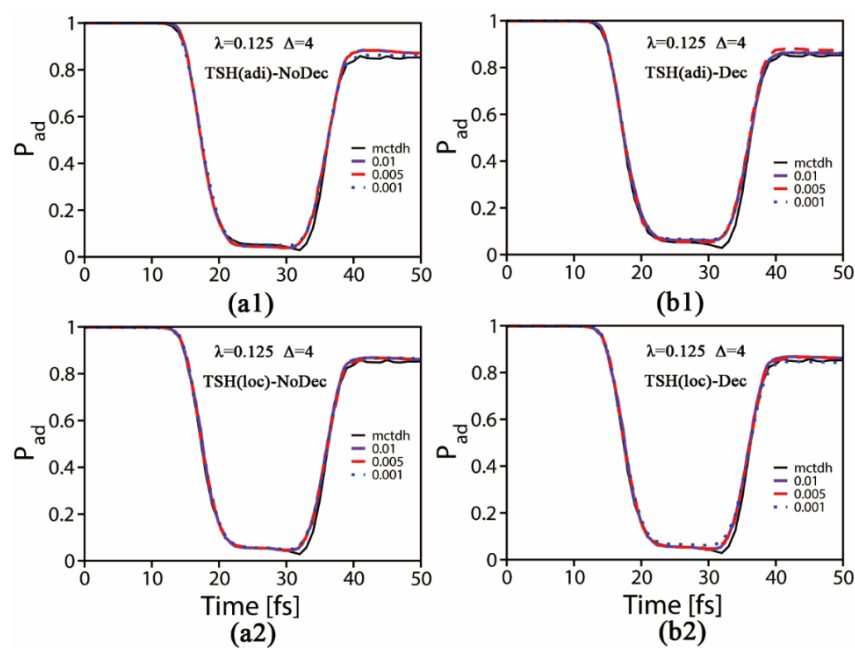